\documentclass[aps,prd,twocolumn,superscriptaddress,10pt,noshowpacs]{revtex4}  
\usepackage[active]{srcltx}
\usepackage{graphicx}
\usepackage[utf8]{inputenc}
\usepackage{amsmath}
\usepackage{times,txfonts}
\usepackage{float}

\begin{document}
\title{Lorentz-violating scalar QED renormalization}

\author{J. Furtado}
\affiliation{Centro de Ci\^{e}ncias e Tecnologia, Universidade Federal do Cariri, 57072-270, Juazeiro do Norte, Cear\'{a}, Brazil}
\email{job.furtado@ufca.edu.br}

\author{R. M. M. Costa Filho}
\affiliation{Universidade Federal de Alagoas - UFAL, 22290-180, Macei\'{o} - AL, Brazil}

\author{J.F. Assun\c{c}\~{a}o}
\affiliation{Universidade Regional do Cariri - URCA, 63180-000, Juazeiro do Norte - CE, Brazil}

\date{\today}

\begin{abstract}
This paper presents divergent contributions of the radiative corrections for a Lorentz-violating extension of the scalar electrodynamics. We initially discuss some features of the model and extract the Feynman rules. Then we compute the one-loop radiative corrections using Feynman parametrization and dimensional regularization in order to evaluate the integrals. We also discuss Furry's theorem validity and renormalization in the present context.
\end{abstract}

\maketitle

\section{Introduction}

The Standard Model (SM) of particle physics, which describes three of the four fundamental interactions, namely the electromagnetic, weak and strong interactions, possesses symmetry $SU(3)\times SU(2)\times U(1)$ and it is invariant under Lorentz and CPT transformations. Despite its phenomenological success the SM leaves some open questions, such as the hierarchy problem and the lack of a quantum description of gravity. Hence it is believed that the SM must be the low energy limit of a more fundamental theory that should unify the SM and the gravitational interaction, providing a quantum description of gravity.

The String Theory \cite{Polchinski} provides a quantum description of gravity, however, it lies on the Planck scale, which is of the order of $10^{19}GeV$, seventeen magnitudes order greater than the electroweak scale associated with the SM. Hence, due to the impossibility of direct experimental verification of string theory an extension of the SM searching for low energy measurable effects was proposed by Kostelecky \cite{Kos1, Kos2, Kos3, Kos4}. The Standard Model Extension (SME) \cite{Colladay:1996iz, Colladay:1998fq} is an effective field theory that includes in its lagrangian all the possible terms that violate Lorentz and CPT symmetries. More recently the SME was divided into two sectors, the minimal sector which possesses only operators with mass dimension $d\leq 4$, and the non-minimal sector with mass dimension operators $d\geq 5$.

The minimal extension of SME was extensively studied in the last years in several contexts, such as radiative corrections \cite{Kos5, Kos6}, neutrinos oscillations \cite{Neut1, Neut2, Neut3}, Euler-Heisenberg effective action \cite{Furtado:2014cja}, gravitational context \cite{Accioly:2016yht, Kostelecky:2016kkn}, among others. The non-minimal extension, despite its posses only non-renormalizable terms, have been receiving attention in the literature \cite{Myers:2003fd, Kostelecky:2009zp, Kostelecky:2011gq, Mariz:2010fm, Rubtsov:2012kb}. A strong motivation lies in the fact that some relevant astrophysical processes impose severe restrictions on the coefficients associated with the operators with dimension $d\geq 5$, being such contributions comparable or even dominant when compared with the ones that arise from dimension $d\leq 4$ operators.

Recently a Lorentz-violating extension of the scalar sector was proposed by Kostelecky \cite{Edwards:2018lsn} in which is presented a general effective scalar field theory in any spacetime dimension that contains explicit perturbative spin-independent Lorentz violating operators of arbitrary mass dimension. This construction is important because the great majority of the fundamental particles of the SM have spin, being the Higgs boson the only example of a fundamental spinless particle in the SM. Despite the minor role played by the scalar sector of QED (sQED), compared to the strong interaction, in describing the coupling between mesons, it was argued \cite{Edwards:2019lfb} that a Lorentz-violating extension of sQED could be an effective way of treating small CPT deviations in neutral-mesons oscillations.

In this paper, we evaluate the divergent contributions from the one-loop radiative corrections for the Lorentz violating scalar extension of the scalar electrodynamics proposed in \cite{Edwards:2018lsn}. We perform the calculations by the explicit computation of all Feynman diagrams involved considering only linear contributions in the Lorentz violating terms. We also calculate all the renormalized parameters besides checking the validity of Furry's theorem.

This paper is organized as follows: in the next section, we present the model itself, some properties regarding the discrete symmetries, and the Feynman rules. In section III, we calculate the one-loop radiative corrections for the photon and scalar propagators, as well as the one-loop corrections for the vertex. In section IV, we present a generalization of the Furry's theorem for the present model, in section V, we renormalize it and, in section VI, we draw our final remarks. 

\section{Model and Feynman rules}
The model we are considering consists of a minimal coupling between the scalar sector of the Lorentz-violating extension of the standard model, recently proposed by Kostelecky \cite{Edwards:2018lsn}, with the electromagnetic field and the usual Maxwell term as the pure gauge sector. Hence the lagrangian describing the system is

\begin{eqnarray}
\nonumber\mathcal{L}&=&G^{\mu\nu}(D_{\mu}\phi)^*D_{\nu}\phi-m^2\phi^*\phi-\frac{i}{2}[\phi^*\hat{k}_a^{\mu}D_{\mu}\phi-\phi\hat{k}_a^{\mu}(D_{\mu}\phi)^*]\\
&&-\frac{1}{4}F_{\mu\nu}F^{\mu\nu}+\frac{\lambda}{4}(\phi^*\phi)^2,
\end{eqnarray}
where the covariant derivative is defined as $D_{\mu}=\partial_{\mu}-ieA_{\mu}$, and the tensor $G^{\mu\nu}=g^{\mu\nu}+(\hat{k}_c)^{\mu\nu}$ is composed by the Minkowski metric tensor $g^{\mu\nu}=diag(1,-1,-1,-1)$ and a Lorentz-violating traceless constant tensor $(\hat{k}_c)^{\mu\nu}$. The tensors $(\hat{k}_c)^{\mu\nu}$ and $\hat{k}_a^{\mu}$ promotes the violation of the Lorentz invariance by breaking the equivalence between particle and observer transformations. Such tensors, assumed to be constant, imply the independence of the space-time position, which yields translational invariance assuring the conservation of momentum and energy. 

In order to extract the Feynman rules of the model, it is convenient to split the Lagrangian into three pieces,

\begin{equation}
    \mathcal{L}=\mathcal{L}_S+\mathcal{L}_G+\mathcal{L}_I,
\end{equation}
being

\begin{eqnarray}
\mathcal{L}_G&=&-\frac{1}{4}F_{\mu\nu}F^{\mu\nu},\\
\mathcal{L}_S&=&G^{\mu\nu}(\partial_{\mu}\phi)^*\partial_{\nu}\phi-m^2\phi^*\phi-\frac{i}{2}[\phi^*\hat{k}_a^{\mu}\partial_{\mu}\phi-\phi\hat{k}_a^{\mu}(\partial_{\mu}\phi)^*],\\
\nonumber\mathcal{L}_I&=&ieG^{\mu\nu}A_{\mu}\phi^*\partial_{\nu}\phi-ieG^{\mu\nu}\partial_{\mu}\phi^*A_{\nu}\phi+e^2G^{\mu\nu}A_{\mu}\phi^*A_{\nu}\phi\\
&&-e\phi^*\hat{k}_a^{\mu}A_{\mu}\phi+\frac{\lambda}{4}(\phi^*\phi)^2.
\end{eqnarray}
From the Gauge sector, we obtain the photon propagator, which is usual. The scalar sector gives us the scalar field propagator. The presence of the Lorentz violating terms gives rise to two insertions in the scalar propagator. A new vertex is constructed from the Lagrangian of interaction, which also has two contributions, one from $(\hat{k}_c)^{\mu\nu}$ and the other from $(\hat{k}_a)^{\mu}$. In summary, the Feynman rules are composed by the usual scalar QED, which are:

\begin{subequations}\label{Feynmanrules1}
\begin{eqnarray}
\raisebox{-0.2cm}{\includegraphics[angle=0,scale=0.3]{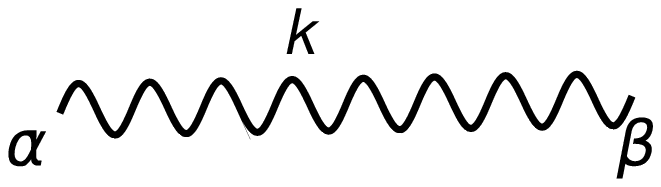}}\label{propf}
 &=& -\frac{ig^{\alpha\beta}}{k^2}=D^{\alpha\beta}(k)\\
 \raisebox{-0.0cm}{\includegraphics[angle=0,scale=0.3]{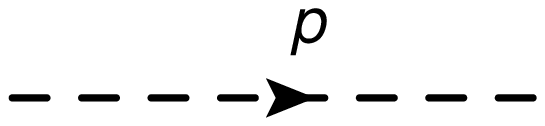}}\label{prop0}
 &=& \frac{i}{p^2-m^2}=S(p)\\
\raisebox{-0.6cm}{\includegraphics[angle=0,scale=0.3]{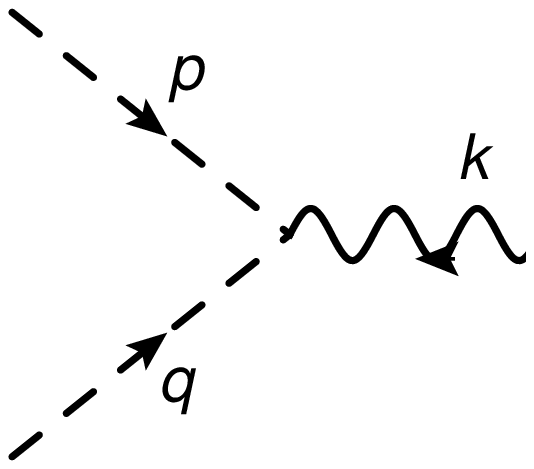}}\label{vert0}
 &=&-ieg^{\beta\mu}(p_{\mu}-q_{\mu})=V_{3}^{\beta}\\
 \raisebox{-0.85cm}{\includegraphics[angle=0,scale=0.3]{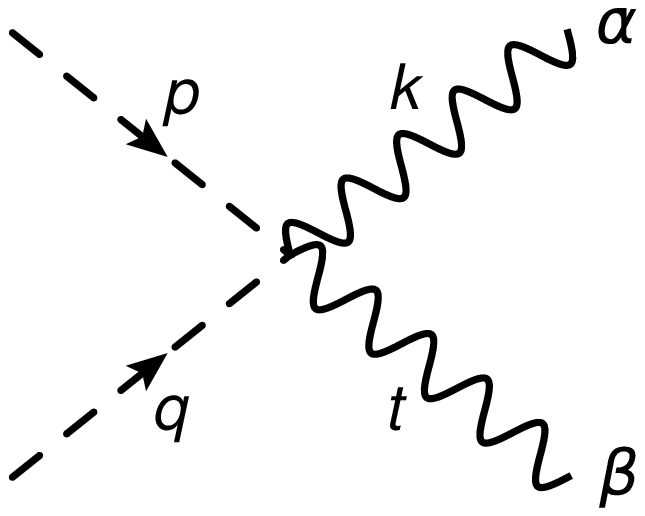}}\label{vert4}
 &=&2ie^2g^{\alpha\beta}=V_{4}^{\alpha\beta},\\
 \raisebox{-0.85cm}{\includegraphics[angle=0,scale=0.3]{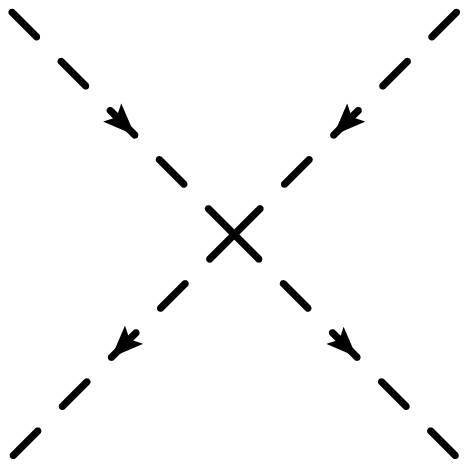}}\label{vert5}
 \hspace{0.5cm}&=&-i\lambda
\end{eqnarray}
\end{subequations}
and the new vertex and propagators insertions, 

\begin{subequations}\label{Feynmanrules2}
\begin{eqnarray}
\raisebox{-0.0cm}{\includegraphics[angle=0,scale=0.3]{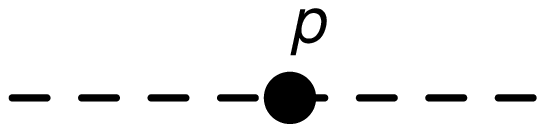}}\label{prop1}
 &=& i(\hat{k}_c)^{\mu\nu}p_{\mu}p_{\nu}\\
 \raisebox{-0.0cm}{\includegraphics[angle=0,scale=0.3]{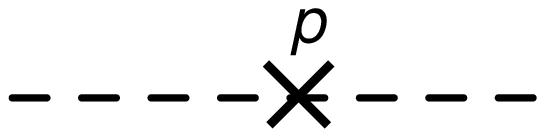}}\label{prop2}
 &=& i\hat{k}_a^{\mu}p_{\mu}\\
\raisebox{-0.60cm}{\includegraphics[angle=0,scale=0.3]{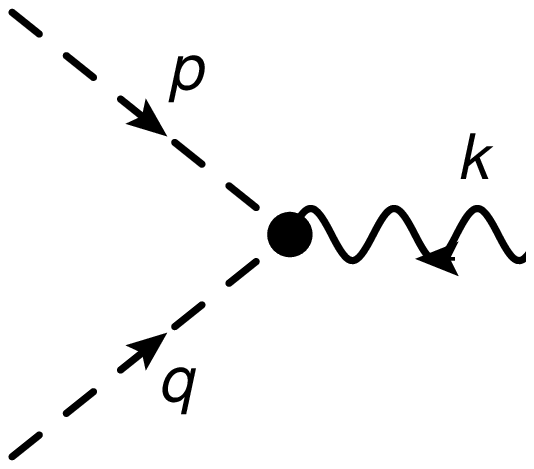}}\label{vert1}
 &=&-ie(\hat{k}_c)^{\beta\mu}(p_{\mu}-q_{\mu})=V_{3kk}^{\beta}\\
 \raisebox{-0.60cm}{\includegraphics[angle=0,scale=0.3]{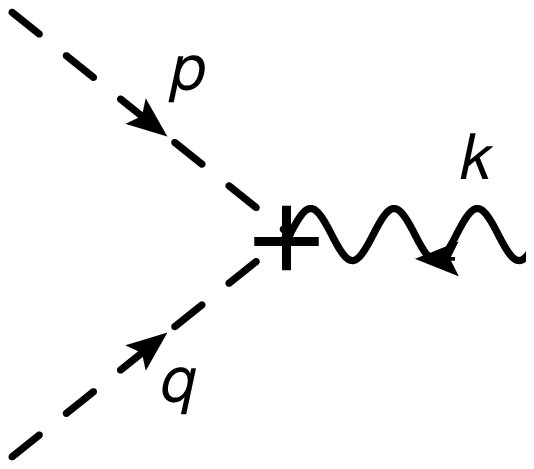}}\label{vert3}
 &=&-ie(\hat{k}_a)^{\beta}=V_{3k}^{\beta}\\
\raisebox{-0.85cm}{\includegraphics[angle=0,scale=0.3]{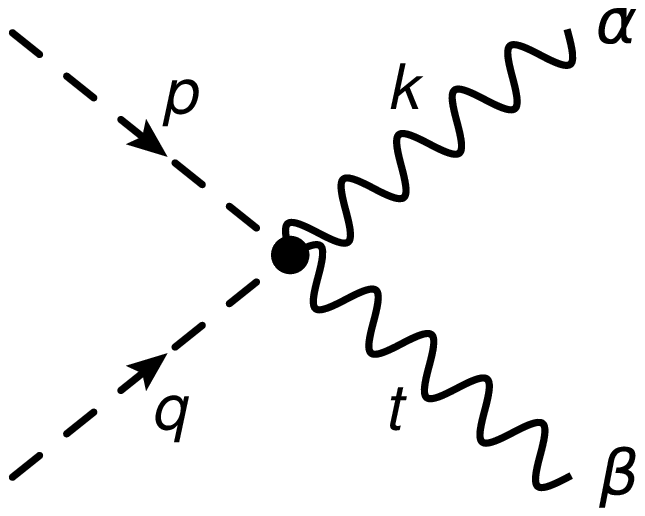}}\label{vert2}
 &=&2ie^2(\hat{k}_c)^{\alpha\beta}=V_{4k}^{\alpha\beta}.
\end{eqnarray}
\end{subequations}

Regarding the analysis of the discrete symmetries on the LV tensors $\hat{k}_a^{\mu}$ and $(\hat{k}_c)^{\mu\nu}$, the results are summarized in the table. As we can see the tensor $\hat{k}_a^{\mu}$ is CPT-odd while $(\hat{k}_c)^{\mu\nu}$ is CPT-even. The PT symmetry is always preserved, however, effects of CP violation can be sawed with the $\hat{k}_a^{0}$ and $(\hat{k}_c)^{0i}$ components. We can also see that there is a violation of the charge conjugation symmetry in every component of the tensor $\hat{k}_a^{\mu}$. Such a violation plays a major role in the calculation of the one-loop radiative corrections, as it will be shown later.

\begin{table}[h!]
\begin{tabular}{|l|l|l|l|l|}
\hline
 & C & P & T & CPT \\ \hline
\,\,\,\,\,\,\,\,\,\,\,\,$\hat{k}_a^{0}$ & - & +  & +  & \,\,\,\,- \\ \hline
\,\,\,\,\,\,\,\,\,\,\,\,$\hat{k}_a^{i}$ & - & -  & -  & \,\,\,\,- \\ \hline
$(\hat{k}_c)^{00}$, $(\hat{k}_c)^{ij}$ & + & + & + & \,\,\,+ \\ \hline
\,\,\,\,\,\,\,\,\,\,$(\hat{k}_c)^{0i}$ & + & - & - & \,\,\,+ \\ \hline
\end{tabular}
\end{table}

\section{Radiative corrections}

In this section, we will calculate the divergent contributions of the one-loop radiative corrections for the present Lorentz violating scalar sector of QED. The calculations were done using standard Feynman diagram computation and dimensional regularization. 

\subsection{Vacuum polarization}

Initially, we will calculate the vacuum polarization considering the coefficient $(\hat{k}_c)^{\mu\nu}$. The complete polarization tensor which gives us the correction for the photon propagator is given by
\begin{equation}
    T^{\mu\nu}(k)=T_a^{\mu\nu}(k)+T_b^{\mu\nu}(k)+T_c^{\mu\nu}(k)+T_d^{\mu\nu}(k)+T_e^{\mu\nu}(k)+T_f^{\mu\nu}(k).
\end{equation}
Each one of the contributions in $T^{\mu\nu}(k)$ stand for the following Feynman graphs

\begin{subequations}\label{scalarloops}
\begin{eqnarray}
\raisebox{-0.85cm}{\includegraphics[angle=0,scale=0.4]{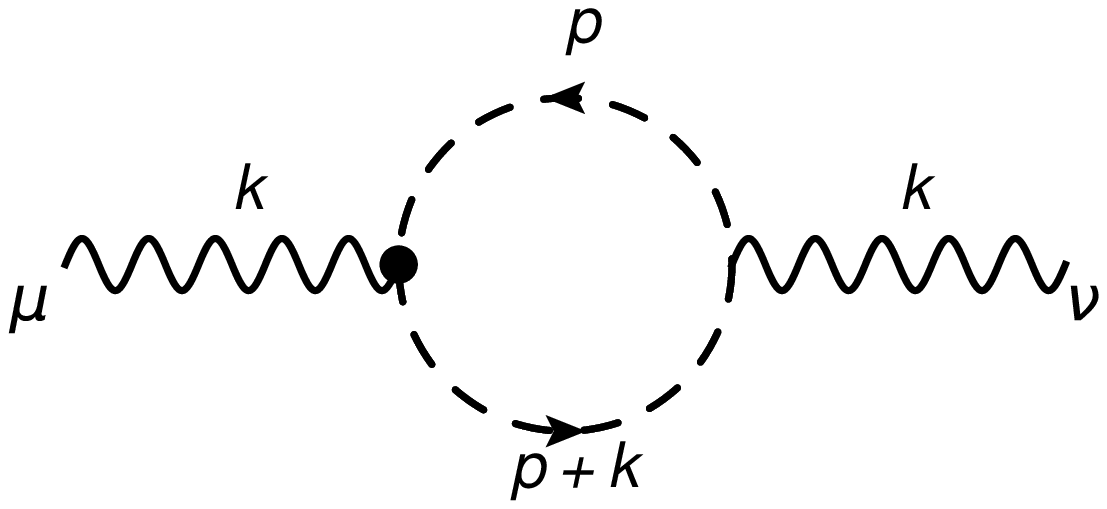}}\label{scalarloop1}
 &=& T_{a}^{\mu\nu}(k)\\
\raisebox{-0.85cm}{\includegraphics[angle=0,scale=0.4]{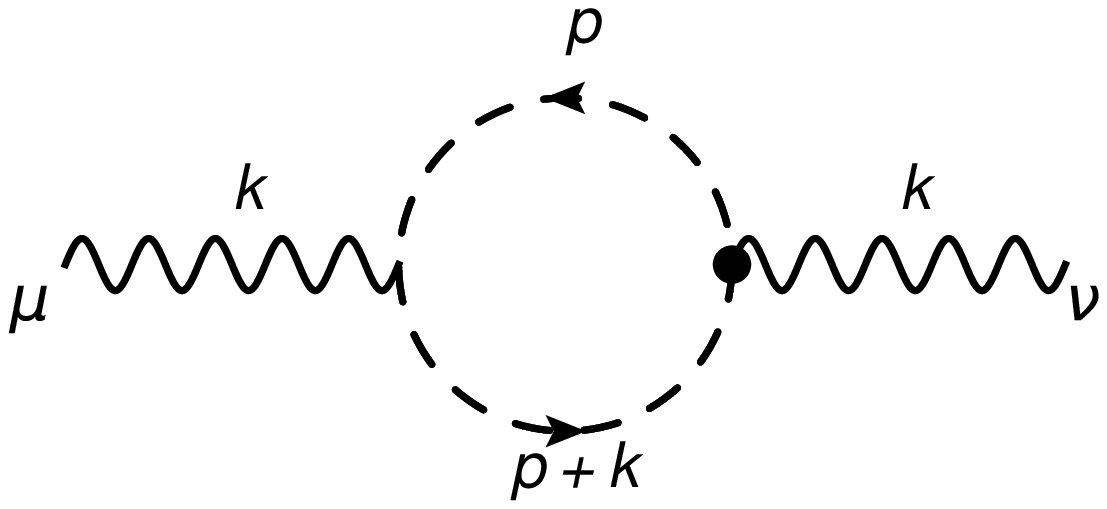}}\label{scalarloop2}
 &=&T_b^{\mu\nu}(k)\\
 \raisebox{-0.85cm}{\includegraphics[angle=0,scale=0.4]{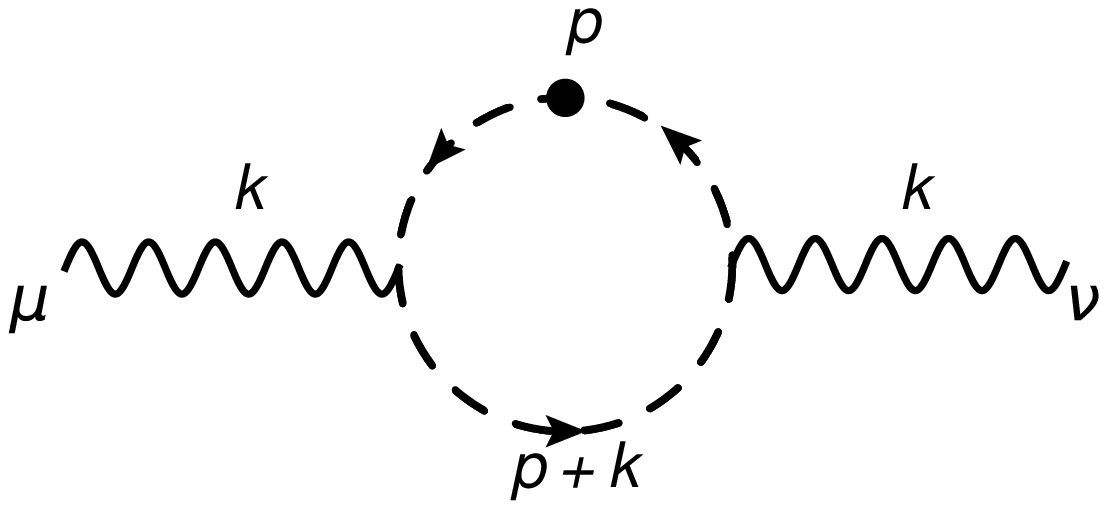}}\label{scalarloop3}
 &=& T_{c}^{\mu\nu}(k)\\
\raisebox{-0.85cm}{\includegraphics[angle=0,scale=0.4]{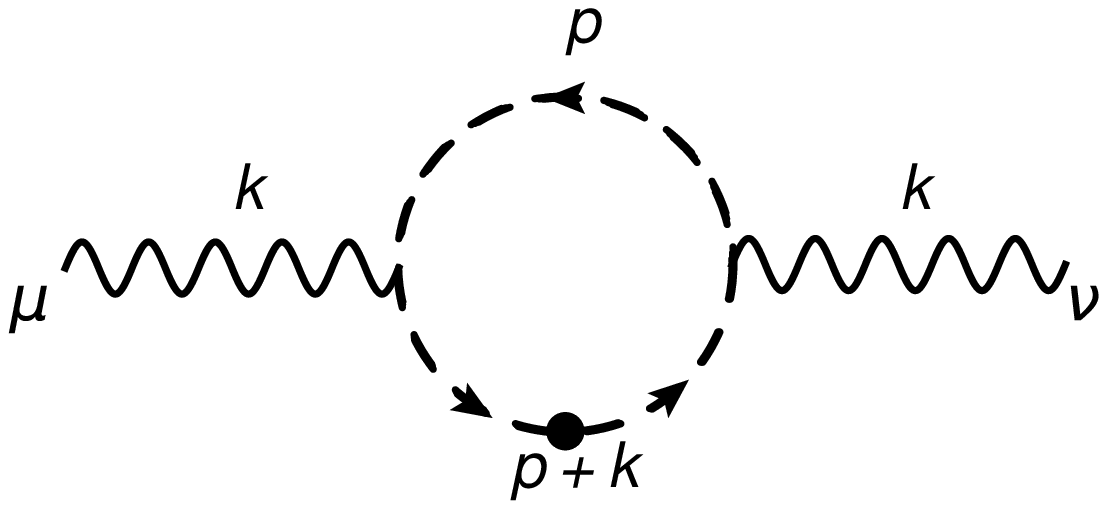}}\label{scalarloop4}
 &=&T_d^{\mu\nu}(k)\\
 \raisebox{-0.85cm}{\includegraphics[angle=0,scale=0.4]{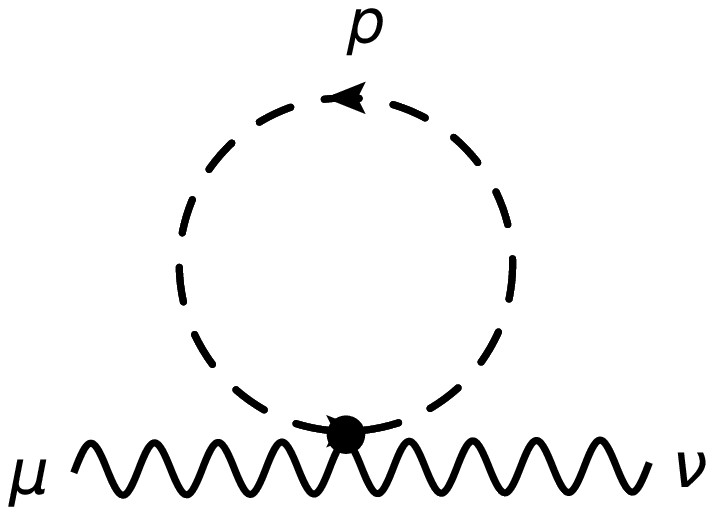}}\label{tadpole1}
 \hspace{0.7cm}&=&T_e^{\mu\nu}(k)\\
 \raisebox{-0.85cm}{\includegraphics[angle=0,scale=0.4]{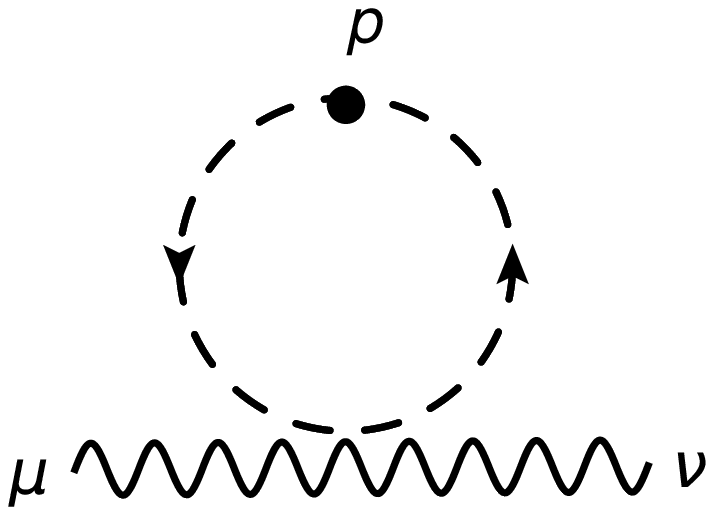}}\label{tadpole2}
 \hspace{0.7cm}&=&T_f^{\mu\nu}(k).
\end{eqnarray}
\end{subequations}
We are considering only the first-order contribution in the coefficient $(\hat{k}_c)^{\mu\nu}$ since it is expected that the bounds on Lorentz violating coefficients to be small. The integrals related to each one of the above graphs are the following:

\begin{eqnarray}
T_a^{\mu\nu}(k)&=&\int\frac{d^4p}{(2\pi)^4}S(p)V_{3kk}^{\mu}S(p_1)V_3^{\nu},\\
T_b^{\mu\nu}(k)&=&\int\frac{d^4p}{(2\pi)^4}S(p)V_{3}^{\mu}S(p_1)V_{3kk}^{\nu},\\
T_c^{\mu\nu}(k)&=&\int\frac{d^4p}{(2\pi)^4}S(p)[i(\hat{k}_c)^{\alpha\beta}p_{\alpha}p_{\beta}]S(p)V_{3}^{\mu}S(p_1)V_{3}^{\nu},\\
T_d^{\mu\nu}(k)&=&\int\frac{d^4p}{(2\pi)^4}S(p)V_{3}^{\mu}S(p_1)\nonumber\\&&\ \ \ \ \ \ \ \ \ \ \ \ \ \ \times [i(\hat{k}_c)^{\alpha\beta}p_{1\alpha}p_{1\beta}]S(p_1)V_{3}^{\nu},\\
T_e^{\mu\nu}(k)&=&\int\frac{d^4p}{(2\pi)^4}V_{4k}^{\mu\nu}S(p),\\
T_f^{\mu\nu}(k)&=&\int\frac{d^4p}{(2\pi)^4}V_{4}^{\mu\nu}S(p)[i(\hat{k}_c)^{\alpha\beta}p_{\alpha}p_{\beta}]S(p).
\end{eqnarray}
In the above equations $p_1=p+k$. To solve the Feynman graphs properly we make use of Feynman parametrization in order to write the denominators in a more suitable way. Hence, after the Feynman parametrization, we obtain
\begin{eqnarray}
\nonumber T_a^{\mu\nu}(k)&=&e^2(\hat{k}_c)^{\mu\alpha}\int_0^1dx\int\frac{d^4p}{(2\pi)^4}\frac{(2q+k)_{\alpha}(2q+k)^{\nu}}{(p^2-M^2)^2},\\
\nonumber T_b^{\mu\nu}(k)&=&e^2(\hat{k}_c)^{\nu\alpha}\int_0^1dx\int\frac{d^4p}{(2\pi)^4}\frac{(2q+k)^{\mu}(2q+k)_{\alpha}}{(p^2-M^2)^2},\\
\nonumber T_c^{\mu\nu}(k)&=&-2e^2(\hat{k}_c)^{\alpha\beta}\int_0^1dx\int\frac{d^4p}{(2\pi)^4}\frac{xq_{\alpha}q_{\beta}(2q+k)^{\mu}(2q+k)^{\nu}}{(p^2-M^2)^3},\\
\nonumber T_d^{\mu\nu}(k)&=&-2e^2(\hat{k}_c)^{\alpha\beta}\int_0^1dx\int\frac{d^4p}{(2\pi)^4}\frac{(1-x)q_{1\alpha}q_{1\beta}(2q+k)^{\mu}(2q+k)^{\nu}}{(p^2-M^2)^3},\\
\nonumber T_e^{\mu\nu}(k)&=&2e^2(\hat{k}_c)^{\mu\nu}\int\frac{d^4p}{(2\pi)^4}\frac{1}{(p^2-m^2)},\\
\nonumber T_f^{\mu\nu}(k)&=&-2e^2g^{\mu\nu}(\hat{k}_c)^{\alpha\beta}\int\frac{d^4p}{(2\pi)^4}\frac{p_{\alpha}p_{\beta}}{(p^2-m^2)^2},
\end{eqnarray}
where $q_1=q+k$ while
\begin{eqnarray}\label{FP1}
q&=&p-(1-x)k\\
M^2&=&m^2+x(x-1)k^2.
\end{eqnarray}
Note that the use of Feynman parametrization is not required to solve the tadpoles. We employ dimensional regularization in order to isolate the divergencies emergent from the graphs. The procedure consists in to extend the space-time from 4 to D dimensions, so that integration measure goes from $d^4p/(2\pi)^4$ to $\mu^{4-D}d^Dp/(2\pi)^D$, where $\mu$ is a mass regulator. This gives us the following result

\begin{eqnarray}
\nonumber T^{\mu\nu}(k)&=&-\frac{ie^2}{48\pi^2\epsilon}\left\{g^{\mu\nu}[2k_{\alpha}k_{\beta}(\hat{k}_c)^{\alpha\beta}-\kappa k^2]+2k^2(\hat{k}_c)^{\mu\nu}\right.\\
   &&+\left.\kappa k^{\mu}k^{\nu}-2k^{\nu}k_{\alpha}(\hat{k}_c)^{\alpha\mu}-2k^{\mu}k_{\alpha}(\hat{k}_c)^{\alpha\nu} \right\},
\end{eqnarray}
being $\kappa=g^{\alpha\beta}(\hat{k}_{c})_{\alpha\beta}$ and $\epsilon=D-4$. Note that the above expression is gauge invariant. Considering the fact that the tensor $(\hat{k}_c)_{\alpha\beta}$ is traceless, the above expression resumes to

\begin{eqnarray}
    \nonumber T^{\mu\nu}(k)&=&-\frac{ie^2}{24\pi^2\epsilon}\left\{k_{\alpha}k_{\beta}(\hat{k}_c)^{\alpha\beta}g^{\mu\nu}+k^2(\hat{k}_c)^{\mu\nu}\right.\\
   &&-\left.k^{\nu}k_{\alpha}(\hat{k}_c)^{\alpha\mu}-k^{\mu}k_{\alpha}(\hat{k}_c)^{\alpha\nu} \right\}.
\end{eqnarray}

The contributions of the Lorentz-violating tensor $\hat{k}_a^{\mu}$ in the photon propagator are given by the following Feynman graphs

\begin{subequations}\label{scalarloopsX}
\begin{eqnarray}
\raisebox{-0.85cm}{\includegraphics[angle=0,scale=0.4]{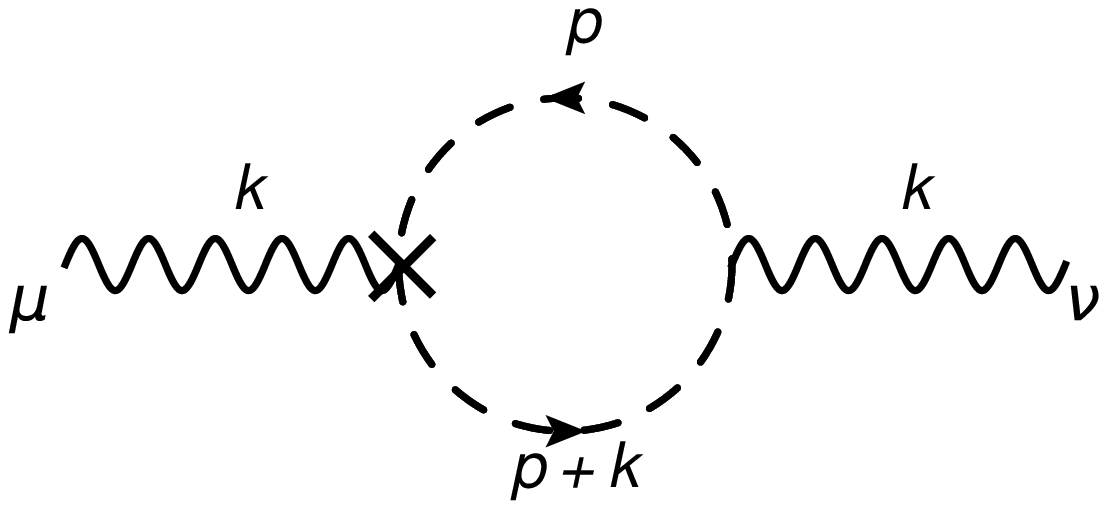}}\label{scalarloopX1}
 &=& R_{a}^{\mu\nu}(k)\\
\raisebox{-0.85cm}{\includegraphics[angle=0,scale=0.4]{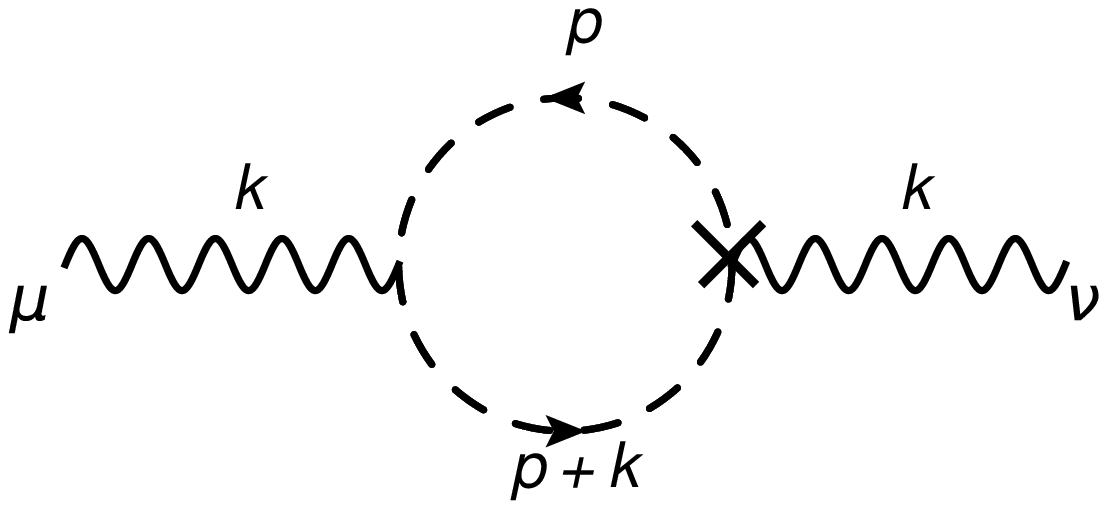}}\label{scalarloopX2}
 &=&R_b^{\mu\nu}(k)\\
 \raisebox{-0.85cm}{\includegraphics[angle=0,scale=0.4]{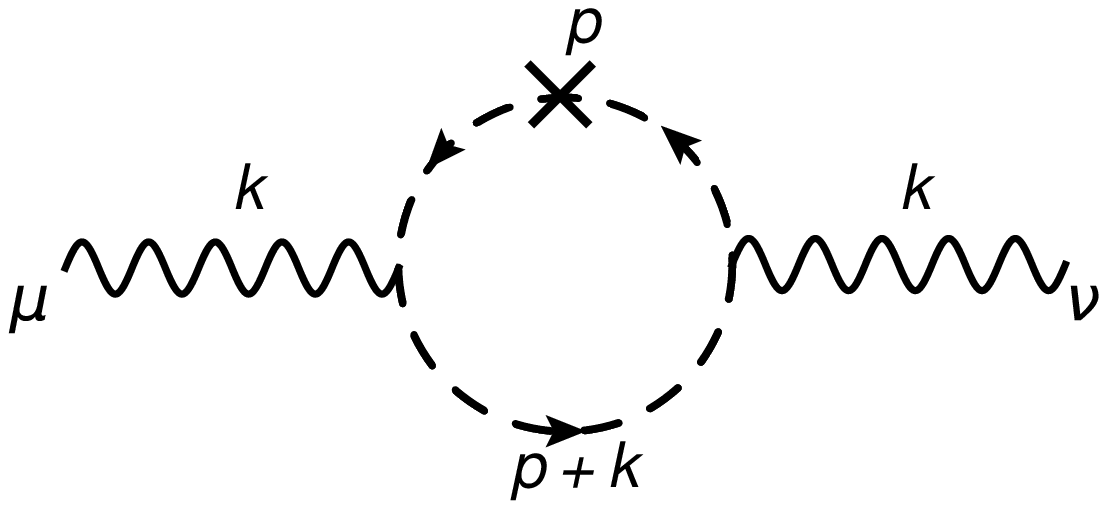}}\label{scalarloopX3}
 &=& R_{c}^{\mu\nu}(k)\\
\raisebox{-0.85cm}{\includegraphics[angle=0,scale=0.4]{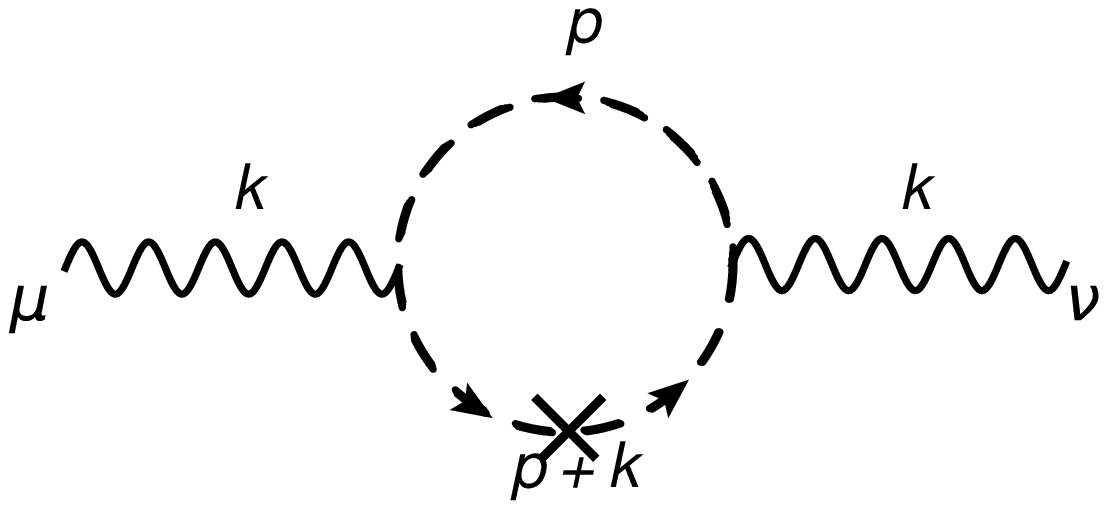}}\label{scalarloopX4}
 &=&R_d^{\mu\nu}(k)\\
  \raisebox{-0.85cm}{\includegraphics[angle=0,scale=0.4]{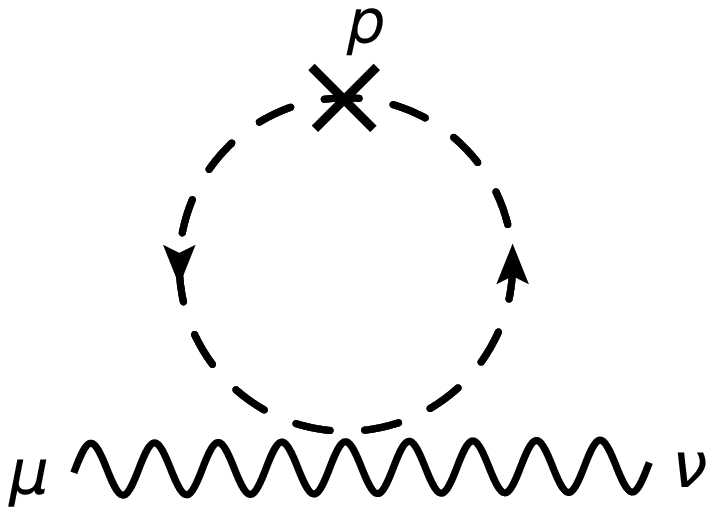}}\label{tadpoleX}
 \hspace{0.7cm}&=&R_e^{\mu\nu}(k).
\end{eqnarray}
\end{subequations}

Analogously to the contribution from $\hat{k}_c^{\mu\nu}$, the contribution from $\hat{k}_a^{\mu}$ is given by the sum of all the graphs in (\ref{scalarloopsX}). Then, we have:
\begin{equation}
    R^{\mu\nu}=R_a^{\mu\nu}+R_b^{\mu\nu}+R_c^{\mu\nu}+R_d^{\mu\nu}+R_e^{\mu\nu}.
\end{equation}
Note that due to the absence of a vertex with two-photon leg there is one less tadpole contribution. All the contributions in the above expression are explicitly written as:
\begin{eqnarray}
R_a^{\mu\nu}(k)&=&\int\frac{d^4p}{(2\pi)^4}S(p)V_{3k}^{\mu}S(p_1)V_3^{\nu},\\
R_b^{\mu\nu}(k)&=&\int\frac{d^4p}{(2\pi)^4}S(p)V_{3}^{\mu}S(p_1)V_{3k}^{\nu},\\
R_c^{\mu\nu}(k)&=&\int\frac{d^4p}{(2\pi)^4}S(p)[i(\hat{k}_a)^{\alpha}p_{\alpha}]S(p)V_{3}^{\mu}S(p_1)V_{3}^{\nu},\\
R_d^{\mu\nu}(k)&=&\int\frac{d^4p}{(2\pi)^4}S(p)V_{3}^{\mu}S(p_1)[i(\hat{k}_a)^{\alpha}p_{1\alpha}]S(p_1)V_{3}^{\nu},\\
R_f^{\mu\nu}(k)&=&\int\frac{d^4p}{(2\pi)^4}V_{4}^{\mu\nu}S(p)[i(\hat{k}_a)^{\alpha}p_{\alpha}]S(p).
\end{eqnarray}
The Feynman parametrization used to solve the above integrals are the same presented in (\ref{FP1}). After employing dimensional regularization we reach at a null result for all the contributions emergent from the tensor $(\hat{k}_a)^{\mu}$, i.e., $R^{\mu\nu}=0$. Such null result can be interpreted as a consequence of the charge symmetry conservation, as we will discuss in more detail in the next section.

\subsection{Scalar propagator correction}

The corrections for the scalar propagator are given by

\begin{equation}
    \Sigma(p)=\Sigma_a(p)+\Sigma_b(p)+\Sigma_c(p)+\Sigma_d(p).
\end{equation}
Analogously to the procedure used to find the corrections for the photon propagator, each term in the above equation is related to one Feynman graph. All the Feynman graphs are depicted below:

\begin{subequations}\label{photonloops}
\begin{eqnarray}
\raisebox{-0.17cm}{\includegraphics[angle=0,scale=0.4]{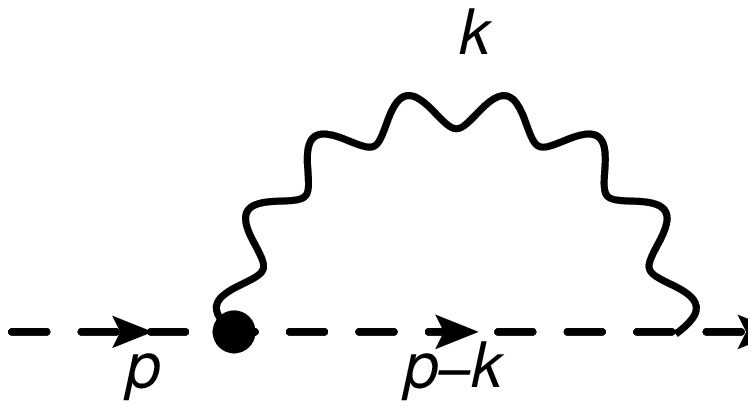}}\label{photonloop1}
 &=& \Sigma_{a}(p)\\
\raisebox{-0.17cm}{\includegraphics[angle=0,scale=0.4]{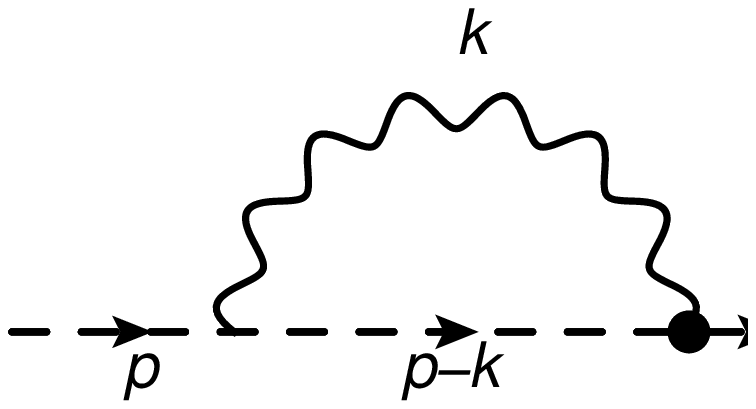}}\label{photonloop2}
 &=&\Sigma_b(p)\\
 \raisebox{-0.17cm}{\includegraphics[angle=0,scale=0.4]{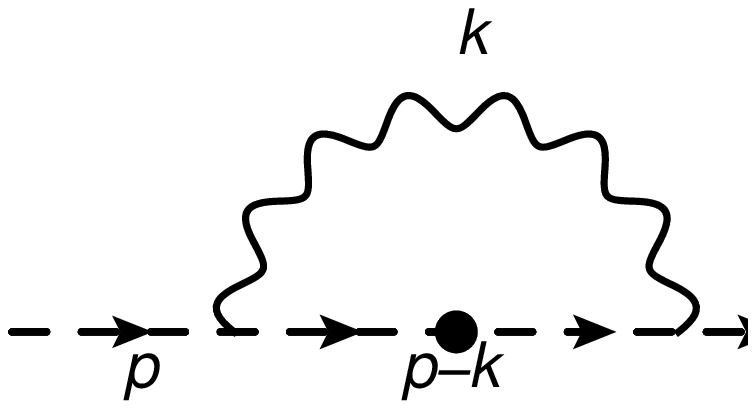}}\label{photonloop3}
 &=&\Sigma_c(p)\\
 \raisebox{-0.17cm}{\includegraphics[angle=0,scale=0.4]{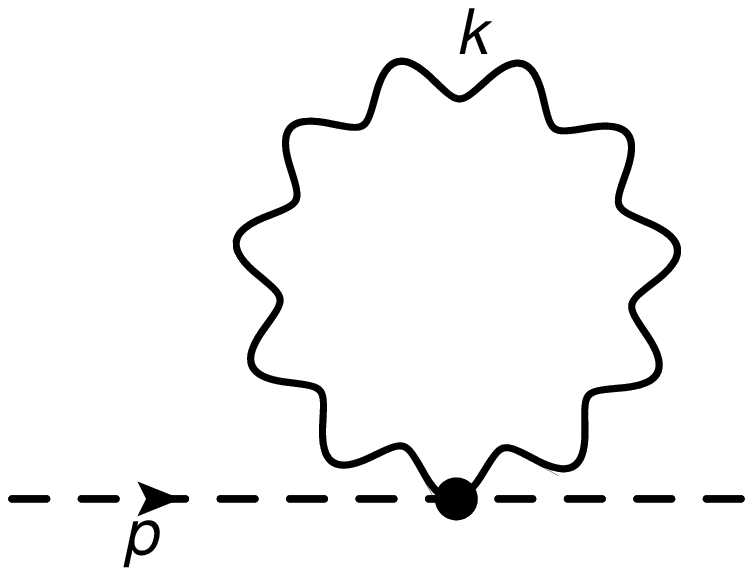}}\label{tadpole3}
 &=&\Sigma_d(p)\\
 \raisebox{-0.10cm}{\includegraphics[angle=0,scale=0.5]{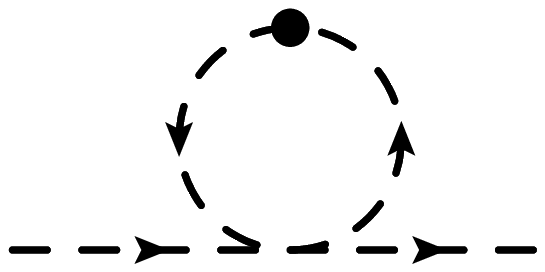}}\label{tadpole4}
 \,\,\,\,\,\,\,&=&\Sigma_e(p)
\end{eqnarray}
\end{subequations}
Where the quantities $\Sigma_a(p)$, $\Sigma_b(p)$, $\Sigma_c(p)$ and $\Sigma_d(p)$ are defined by

\begin{eqnarray}
\Sigma_a(p)&=&\int\frac{d^4 k}{(2\pi)^4}V_{3L}^{\mu}S(p-k)V_3^{\nu}D_{\mu\nu}(k),\\
\Sigma_b(p)&=&\int\frac{d^4 k}{(2\pi)^4}V_{3}^{\mu}S(p-k)V_{3L}^{\nu}D_{\mu\nu}(k),\\
\Sigma_c(p)&=&\int\frac{d^4 k}{(2\pi)^4}V_{3}^{\mu}S(p)[i(\hat{k}_c)^{\alpha\beta}p_{\alpha}p_{\beta}]S(p)V_{L}^{\nu}D_{\mu\nu}(k),\\
\Sigma_d(p)&=&\int\frac{d^4 k}{(2\pi)^4}V_{4k}^{\mu\nu}D_{\mu\nu}(k),\\
\Sigma_e(p)&=&\int\frac{d^4 k}{(2\pi)^4}S(k)[i(\hat{k}_c)^{\alpha\beta}p_{\alpha}p_{\beta}]S(k)(-i\lambda).
\end{eqnarray}
We also make use of Feynman parametrization to write the denominators in a proper way. This gives us the following expressions

\begin{eqnarray}
\nonumber\Sigma_a(p)&=&-e^2(\hat{k}_c)^{\mu\alpha}\int_0^1 dx\int\frac{d^4k}{(2\pi)^4}\frac{(2p-q)_{\alpha}(2p-q)_{\mu}}{(k^2-M^2)^2},\\
\nonumber\Sigma_b(p)&=&-e^2(\hat{k}_c)^{\alpha\nu}\int_0^1 dx\int\frac{d^4k}{(2\pi)^4}\frac{(2p-q)_{\nu}(2p-q)_{\alpha}}{(k^2-M^2)^2},\\
\nonumber\Sigma_c(p)&=&2e^2(\hat{k}_c)^{\alpha\beta}\int_0^1 dx\int\frac{d^4k}{(2\pi)^4}\frac{x(2p-q)_{\mu}q_{1\alpha}q_{1\beta}(2p-q)^{\mu}}{(k^2-M^2)^3},\\
\nonumber\Sigma_d(p)&=&-2e^2(\hat{k}_c)^{\mu}_{\,\,\,\,\mu}\int\frac{d^4k}{(2\pi)^4}\frac{1}{k^2},\\
\nonumber\Sigma_e(p)&=&-\lambda(\hat{k}_c)^{\alpha\beta}\int\frac{d^4k}{(2\pi)^4}\frac{k_{\alpha}k_{\beta}}{(k^2-m^2)^2},
\end{eqnarray}
with the following displacements in the momentum and mass:
\begin{eqnarray}
q&=&k+xp,\\
M^2&=&xm^2-(1-x)xp^2.
\end{eqnarray}
As it was already discussed previously, we are considering the tensor $(\hat{k}_c)^{\mu\nu}$ to be traceless. So that the contributions $\Sigma_d(p)$ and $\Sigma_e(p)$ vanish. In order to solve the integrals, we also use dimensional regularization, which gives us

\begin{equation}
    \Sigma(p)=-\frac{ie^2}{2\pi^2\epsilon}(\hat{k}_c^{\alpha\beta})p_{\alpha}p_{\beta}.
\end{equation}

The contribution of $\hat{k}_a^{\mu}$ to the scalar propagator is given by the following graphs
\begin{subequations}\label{photonloopsX}
\begin{eqnarray}
\raisebox{-0.17cm}{\includegraphics[angle=0,scale=0.4]{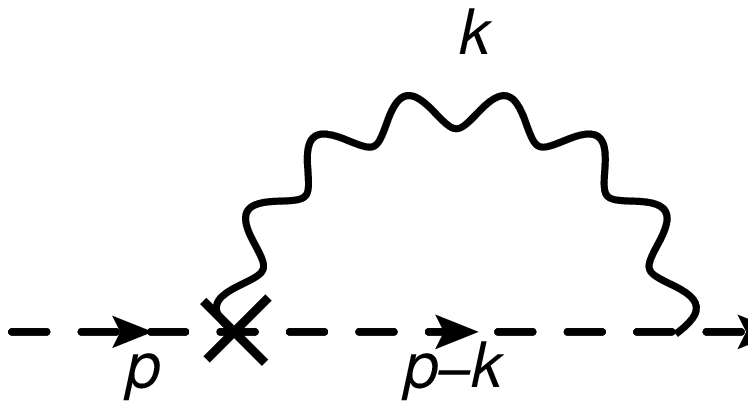}}\label{photonloopX1}
 &=& \Delta_{a}(p)\\
\raisebox{-0.17cm}{\includegraphics[angle=0,scale=0.4]{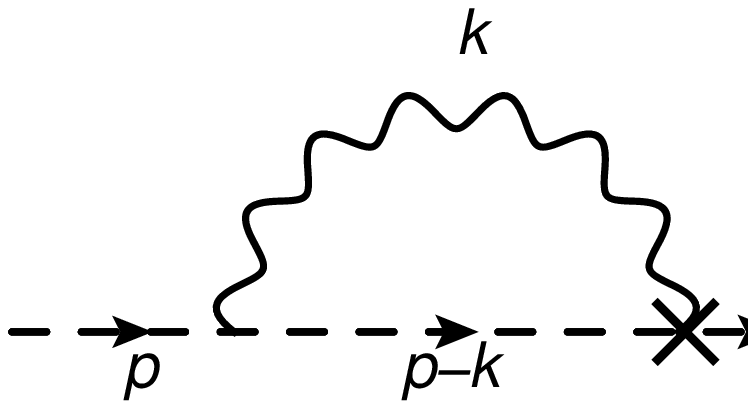}}\label{photonloopX2}
 &=&\Delta_b(p)\\
 \raisebox{-0.17cm}{\includegraphics[angle=0,scale=0.4]{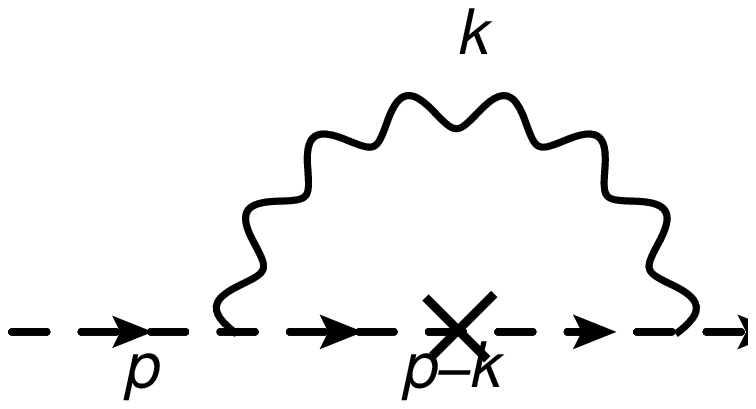}}\label{photonloopX3}
 &=&\Delta_c(p)\\
 \raisebox{-0.10cm}{\includegraphics[angle=0,scale=0.5]{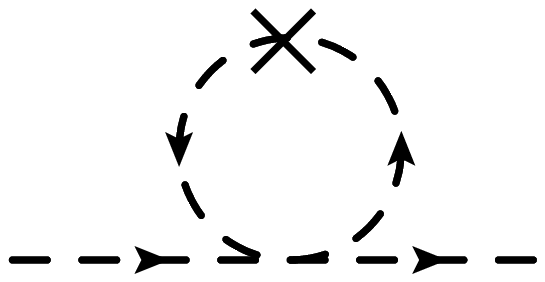}}\label{tadpole5}
 \,\,\,\,\,\,\,&=&\Delta_d(p).
\end{eqnarray}
\end{subequations}
which gives us the following correction for the scalar propagator

\begin{equation}
    \Delta(p)=\Delta_a(p)+\Delta_b(p)+\Delta_c(p)+\Delta_d(p)=-\frac{ie^2}{4\pi^2\epsilon}\hat{k}_a^{\mu}p_{\mu}.
\end{equation}

Note that, unlike the usual scalar QED, there is no $\lambda$ dependence in the corrections to the scalar self-energy. 

\subsection{Scalar-scalar-photon vertex correction}

The vertex corrections are divided into three contributions: those that correct the vertex scalar-scalar-photon, those that correct the vertex scalar-scalar-photon-photon, and those that correct the four-scalar vertex. 

\begin{figure}[ht]
\begin{center}
\includegraphics[scale=0.5]{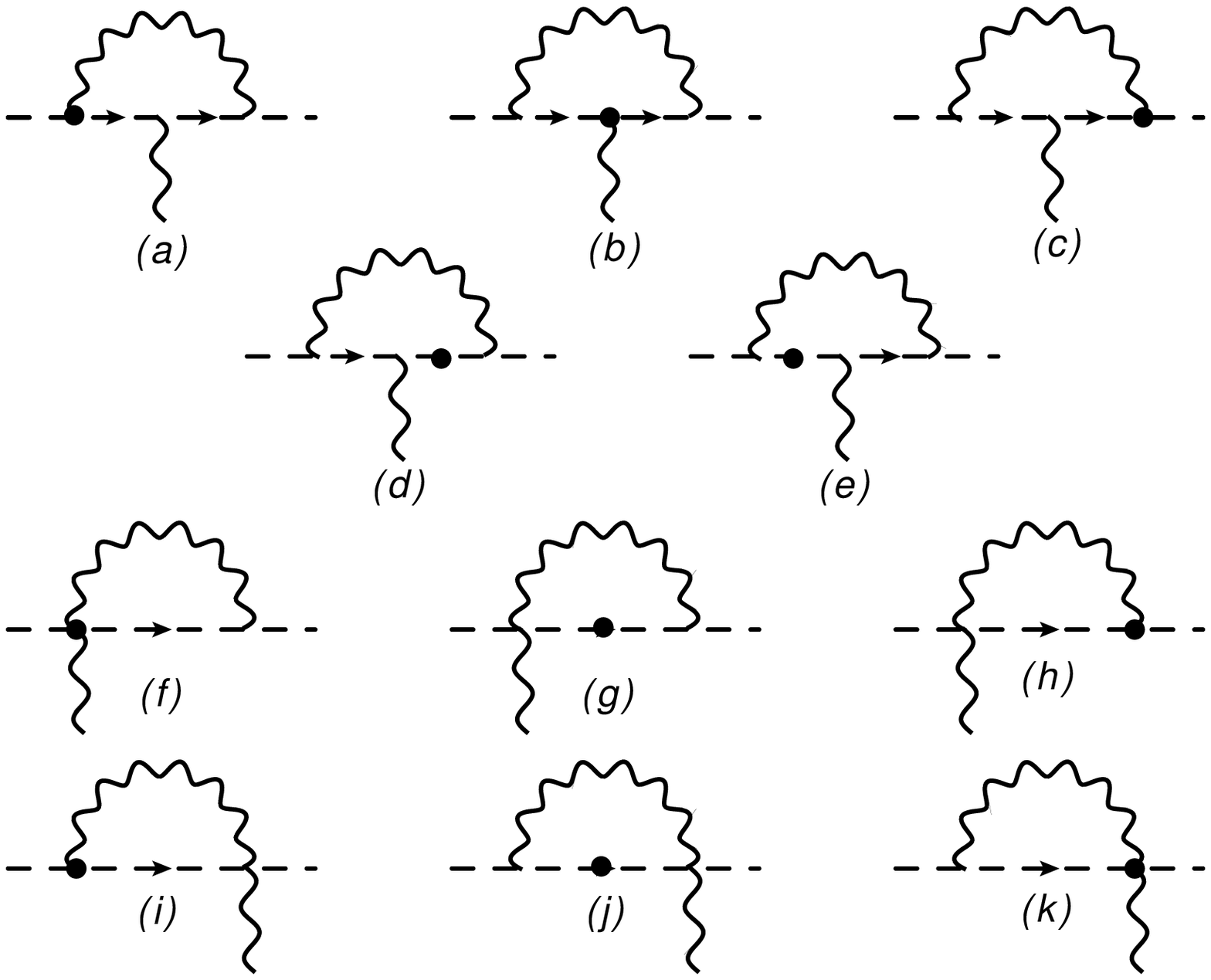}
\caption{Scalar-scalar-photon vertex correction from $(\hat{k}_c)^{\mu\nu}$.}
\label{scalar-scalar-photon1}
\end{center}
\end{figure}

\begin{figure}[ht]
\begin{center}
\includegraphics[scale=0.5]{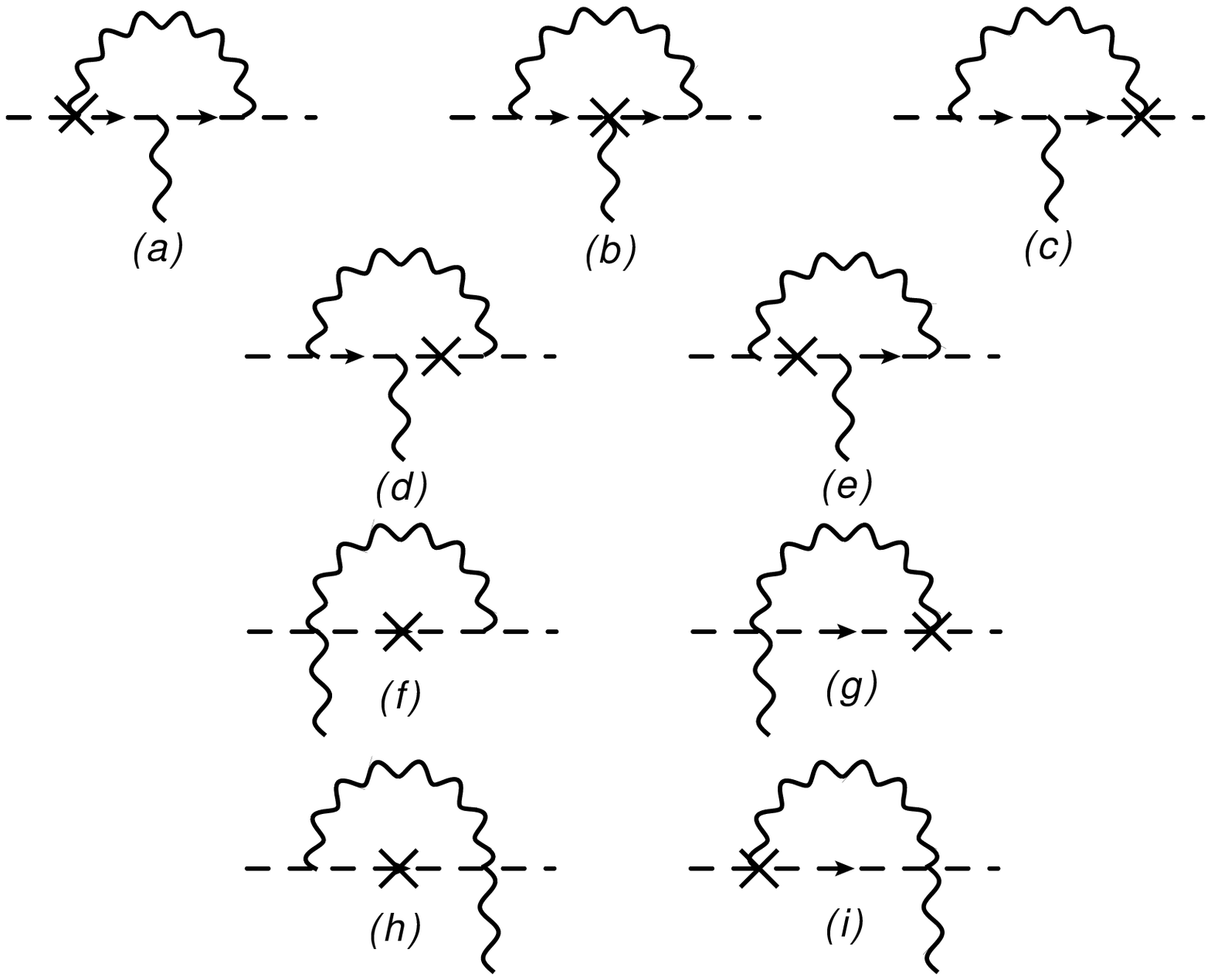}
\caption{Scalar-scalar-photon vertex correction from $(\hat{k}_a)^{\mu}$.}
\label{scalar-scalar-photon2}
\end{center}
\end{figure}

\begin{figure}[H]
 \centering
 \begin{minipage}{.2\textwidth}
   \centering
   \includegraphics[scale=0.45]{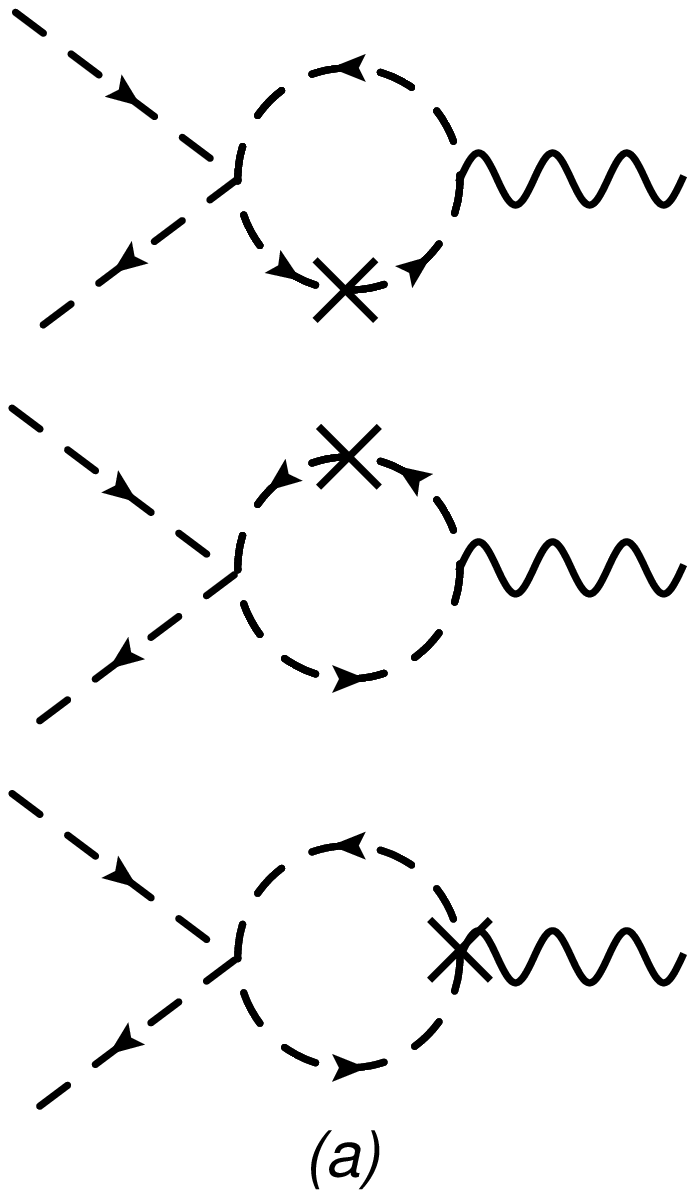}
   \label{fig3a}
 \end{minipage}%
 \begin{minipage}{.2\textwidth}
   \centering
   \includegraphics[scale=0.45]{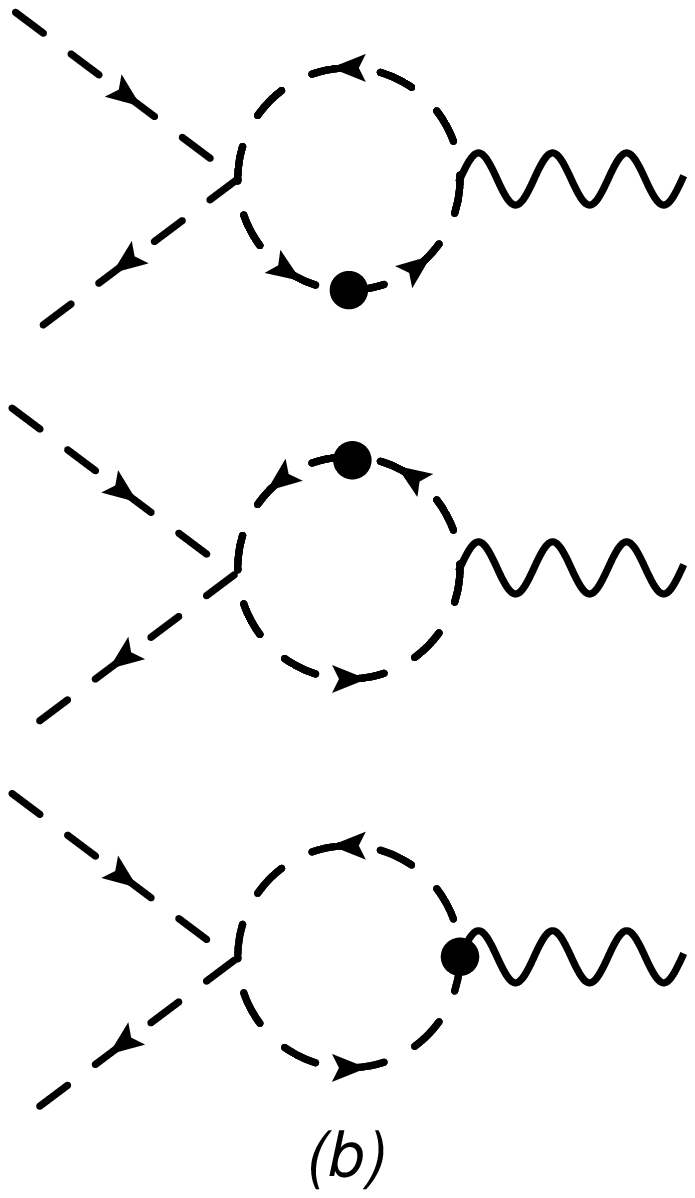}
   \label{fig3b}
 \end{minipage}
 \caption{Scalar-scalar-photon vertex correction with dependence in $\lambda$. Contributions from (a) $(\hat{k}_a)_{\mu}$ and (b) $(\hat{k}_c)_{\mu\nu}$.}
 \label{Fig3}
 \end{figure}

The graphs correcting the scalar-scalar-photon vertex with contribution from the tensor $(\hat{k}_c)^{\mu\nu}$ are depicted in Fig. \ref{scalar-scalar-photon1}, while the corrections emergent from the tensor $(\hat{k}_a)^{\mu}$ are presented in Fig. \ref{scalar-scalar-photon2}, however, the scalar-scalar-photon vertex correction with dependence on $\lambda$ is depicted in Fig. \ref{Fig3}. 

%

All the contributions from Fig. \ref{scalar-scalar-photon1} and Fig. \ref{scalar-scalar-photon2} were explicitly calculated and the results are reported in the equations (\ref{Lambda}). Fig. \ref{Fig3} shows the graphs with corrections to the scalar-scalar-photon vertex with dependence in $\lambda$. Hence the vertex corrections are the following:

\begin{subequations}\label{Lambda}
\begin{eqnarray}
\Lambda_{\mu}^{c}&=&\frac{ie^3}{2\pi^2\epsilon}(\hat{k}_c)_{\mu\alpha}(2p^{\alpha}+p'^{\alpha}),\\
\Lambda_{\mu}^{a}&=&\frac{ie^3}{4\pi^2\epsilon}(\hat{k}_a)_{\mu},
\end{eqnarray}
\end{subequations}
where $p^{\alpha}$ stands for the incoming scalar momentum while $p'^{\alpha}$ stands for the incoming photon momentum. Note that, as expected, there is no $\lambda$ contribution.

\subsection{Scalar-scalar-photon-photon vertex correction}

The divergent contributions for the scalar-scalar-photon-photon correction are then summarized in the following results:

\begin{subequations}
\begin{eqnarray}
\Lambda_{\mu\nu}^c&=&-\frac{ie^4}{\pi^2\epsilon}(\hat{k}_c)_{\mu\nu},\\
\Lambda_{\mu\nu}^a&=&0.
\end{eqnarray}
\end{subequations}

Note that there are no $\lambda$-dependent terms, which is in agreement with Ward's identity.

\subsection{Scalar-scalar-scalar-scalar vertex correction}

Let us, finally, calculate the contributions to the scalar field self-interaction vertex. The contributions which correct the scalar field self-interaction vertex are depicted in the appendix. First, the graphs from (\ref{F9}) vanish. Those with $(\hat{k}_a)^{\mu}$ vanish in the integration level while the ones with $(\hat{k}_c)^{\mu\nu}$ disappear due to the fact that $(\hat{k}_c)^{\mu\nu}$ is traceless. 

The graphs depicted in Fig. \ref{F8} also vanish by the same reason as the ones of Fig. \ref{F9}. The permutations implicit in the figure take into account graphs with opposite charge flow. The Feynman diagrams presented in Fig. \ref{F4} are all finite, as we can see by a simple power counting. In the diagrams depicted in Fig. \ref{F41}, the (a) and (b) contributions are finite, and the (c) and (d) contributions are divergent. However, the divergent term is proportional to $(\hat{k}_c)^{\beta}_{\beta}$, which vanishes since the tensor $(\hat{k}_c)^{\mu\nu}$ is traceless. 
Analogously to the diagrams in Fig. \ref{F4}, the ones in Fig. \ref{F5} are all finite, while those in Fig. \ref{F51} vanish, since the tensor $(\hat{k}_c)^{\mu\nu}$ is traceless. All the diagrams depicted in Fig. \ref{F1} and Fig. \ref{F2} are finite and the divergent contributions of Fig. \ref{F11} and Fig. \ref{F21} are proportional to the trace of $(\hat{k}_c)^{\mu\nu}$.

From a loop of photons, the allowed possibilities for the scalar-scalar scattering are depicted in Fig. \ref{F7}. Note that there is no contribution from $(\hat{k}_a)^{\mu}$ and the contributions from $(\hat{k}_c)^{\mu\nu}$ vanish. 

\section{generalized Furry theorem}

\begin{figure}[h!]
\includegraphics[scale=0.4]{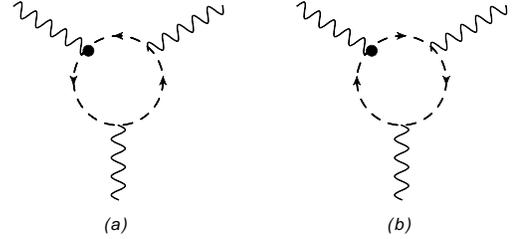}
\caption{Feynman graphs for a scalar loop with three external photon legs with opposite loop orientations - LV vertex.}
\label{Furry1}
\end{figure}

In the spinor QED, the Furry theorem states that any fermion loop with an odd number of external photon legs has a null amplitude. This result is an immediate consequence of the charge conjugation symmetry. In this section, we will establish the validity of Furry's theorem for the Lorentz-violating scalar QED. Initially consider the graphs depicted in Fig. \ref{Furry1}.

\begin{figure}[h!]
\begin{center}
\includegraphics[scale=0.4]{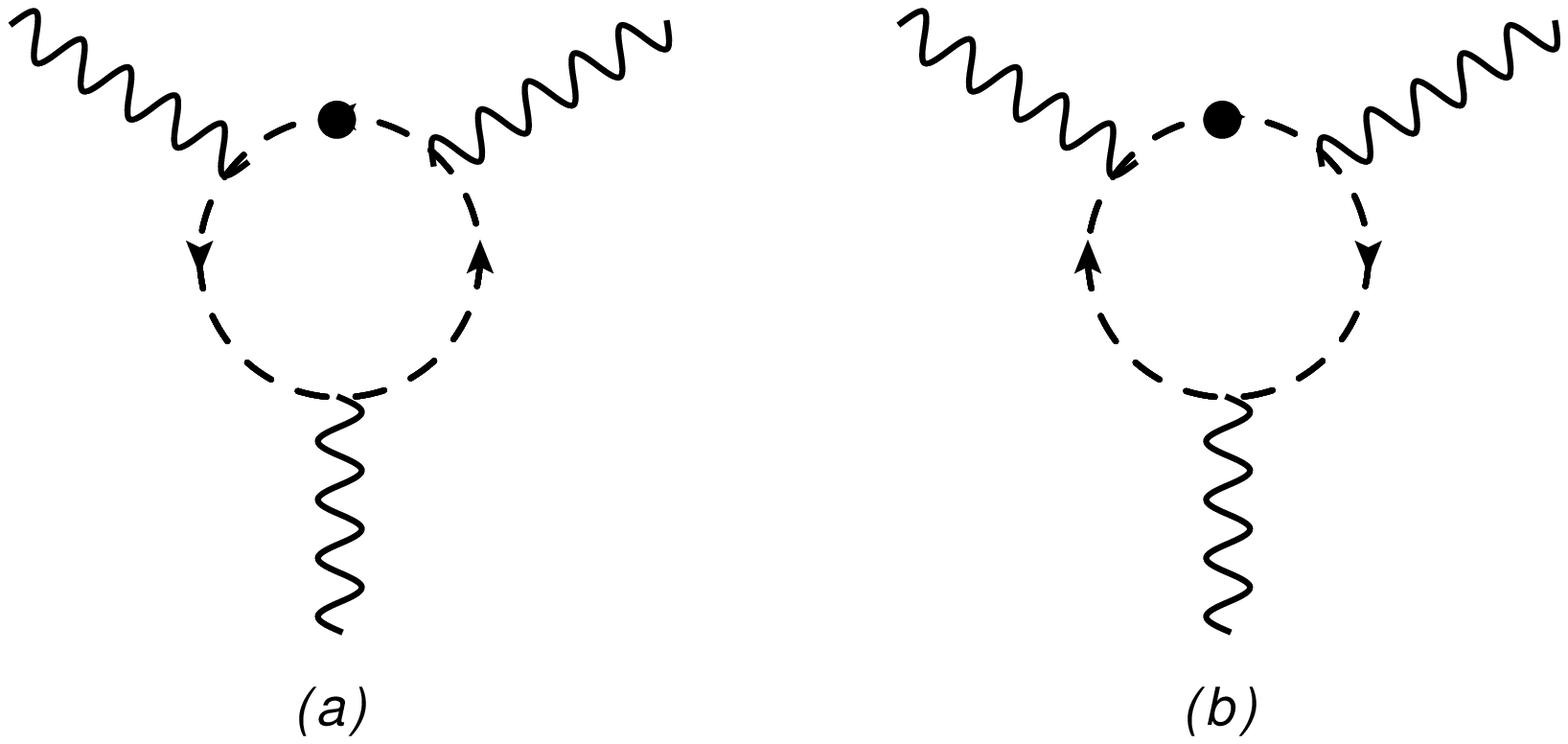}
\caption{Feynman graphs for a scalar loop with three external photon legs with opposite loop orientations - insertion in the propagator.}
\label{Furry2}
\end{center}
\end{figure}

\begin{figure}[h!]
\begin{center}
\includegraphics[scale=0.4]{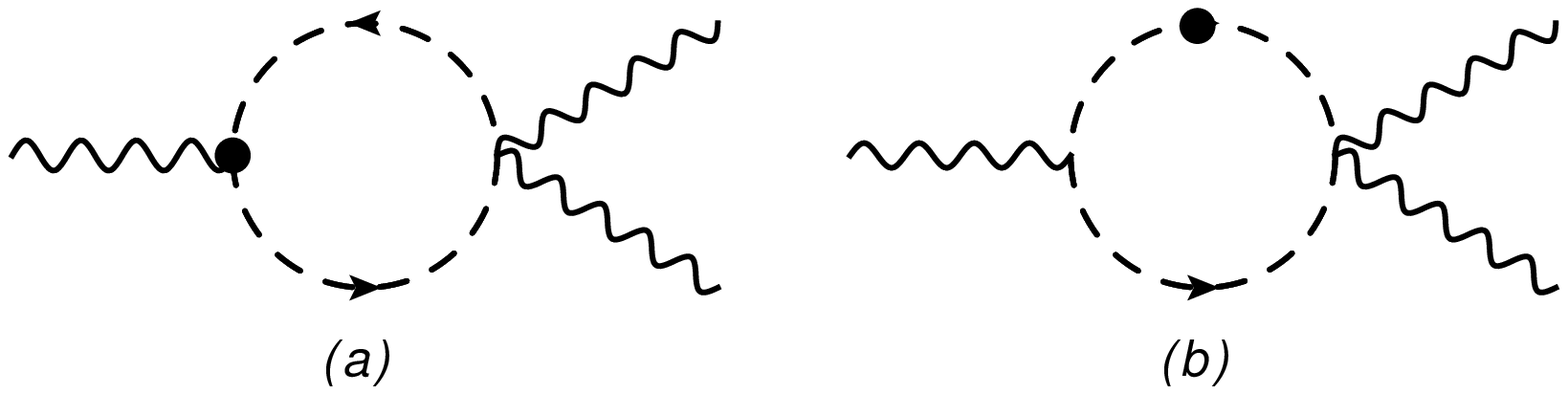}
\caption{Feynman graphs for a scalar loop with three external photon legs with opposite loop orientations - insertion in the propagator.}
\label{Furry3}
\end{center}
\end{figure}

The loop described in Fig. \ref{Furry1}a is written as
\begin{equation}
    L_a^{\mu\nu\lambda}=-e^3\int\frac{d^4p}{(2\pi)^4}\frac{(\hat{k}_c)^{\mu\alpha}(p+p_1)_{\alpha}(p_1+p_{12})^{\nu}(p_{12}+p)^{\lambda}}{(p^2-m^2)(p_1^2-m^2)(p_{12}^2-m^2)},
\end{equation}
while the counterclockwise one, described in Fig. \ref{Furry1}b, is given by
\begin{equation}
    L_b^{\mu\nu\lambda}=-e^3\int\frac{d^4q}{(2\pi)^4}\frac{(q+q_3)^{\lambda}(q_3+q_{32})^{\nu}(q_{32}+q)_{\alpha}(\hat{k}_c)^{\alpha\mu}}{(q^2-m^2)(q_3^2-m^2)(q_{32}^2-m^2)}.
\end{equation}
Being the orientation the only difference between the two loops, it is reasonable to set $q=-p$. Besides, the momentum conservation requires that $\sum k_i=0$, so that, $k_3=-k_1-k_2$. Hence $q_3=-p_{12}$ and $q_{32}=-p_1$. Therefore, it is straightforward to see that
\begin{equation}
    L_{a}^{\mu\nu\lambda}=-L_{b}^{\mu\nu\lambda},
\end{equation}
which means that the total amplitude is null because the contributions with opposite orientations cancel each other. An analogous procedure for the insertion in the propagator, as depicted in Fig. \ref{Furry2}, yields the same result, i.e., the amplitudes with opposite orientations cancel each other.
The same cancellation is evident when we consider the other possible Feynman graphs containing a scalar loop with three external photon legs as depicted in Fig. \ref{Furry3}

Besides that, considering now the contribution emergent from tensor $\hat{k}_a^{\mu}$ it is straightforward to see that the same does not occur. Instead of the cancellation of the opposite orientation contributions, they add to each other. This no cancellation result is a consequence of the charge violation symmetry imposed by the $\hat{k}_a^{\mu}$ tensor. In fact, it is possible to extend the above result of the three external photon legs contribution to any odd number of external photon legs by considering the charge conjugation operator $\hat{C}$. The effect of the charge conjugation operator is not related to the spacetime but to the fields. Such operator act by interchanging particles and antiparticles, so that $\hat{C}\phi(x,t)\hat{C}=\phi^*(x,t)$. However, the net effect of $\hat{C}$ in the Feynman rules can be described by considering $\hat{C}p_{\mu}\hat{C}=-p_{\mu}$, which gives us 

\begin{eqnarray}
\hat{C}S(p)\hat{C}&=&S(-p)=S(p)\\
\hat{C}I_{kk}(p)\hat{C}&=&I_{kk}(-p)=I_{kk}(p)\\
\hat{C}I_{k}(p)\hat{C}&=&I_{k}(-p)=-I_{k}(p)\\
\hat{C}V_3^{\beta}(p,q)\hat{C}&=&V_3^{\beta}(q,p)=-V_3^{\beta}(p,q)\\
\hat{C}V_4^{\alpha\beta}\hat{C}&=&V_4^{\alpha\beta}\\
\hat{C}V_{3kk}^{\beta}(p,q)\hat{C}&=&V_{3kk}^{\beta}(q,p)=-V_{3kk}^{\beta}(p,q)\\
\hat{C}V_{3k}^{\beta}\hat{C}&=&V_{3k}^{\beta}\\
\hat{C}V_{4k}^{\alpha\beta}\hat{C}&=&V_{4k}^{\alpha\beta}.
\end{eqnarray}
Hence, inserting an identity operator identified as $I=\hat{C}\hat{C}$ between each propagator and vertex, it is possible to see the cancellation of any loop with an odd number of external photon legs for the contributions with $(\hat{k}_c)^{\mu\nu}$, and also the cancellation of any loop with an even number of external photon legs for $\hat{k}_a^{\mu}$. An immediate consequence of this fact is the absence of contribution from $\hat{k}_a^{\mu}$ to the Euler-Heisenberg action in scalar QED. 

\section{renormalization}

In this section, we will perform the renormalization Lorentz violating model of scalar electrodynamics. Let us first summarize all the radiative corrections calculated in the previous sections. Considering $\Pi^{\mu\nu}(k)=T^{\mu\nu}(k)+R^{\mu\nu}(k)$, $\Sigma_T(p)=\Sigma(p)+\Delta(p)$, $\Lambda_{\mu}=\Lambda_{\mu}^c+\Lambda_{\mu}^a$ and $\Lambda_{\mu\nu}=\Lambda_{\mu\nu}^c+\Lambda_{\mu\nu}^a$, we have,

\begin{eqnarray}
 \nonumber \Pi^{\mu\nu}(k)&=&-\frac{ie^2}{24\pi^2\epsilon}\left\{k_{\alpha}k_{\beta}(\hat{k}_c)^{\alpha\beta}g^{\mu\nu}+k^2(\hat{k}_c)^{\mu\nu}\right.\\
   &&-\left.k^{\nu}k_{\alpha}(\hat{k}_c)^{\alpha\mu}-k^{\mu}k_{\alpha}(\hat{k}_c)^{\alpha\nu} \right\},\\
   \Sigma_T(p)&=&-\frac{ie^2}{4\pi^2\epsilon}\left[(\hat{k}_a)^{\mu}p_{\mu}+2(\hat{k}_c)_{\mu\nu}p^{\mu}p^{\nu}\right],\\
   \Lambda_{\mu}&=&\frac{ie^3}{4\pi^2\epsilon}\left[4p^{\alpha}(\hat{k}_c)_{\alpha\mu}+(\hat{k}_a)_{\mu}\right],\\
   \Lambda_{\mu\nu}&=&\frac{ie^2}{\pi^2\epsilon}(\hat{k}_c)_{\mu\nu}.
\end{eqnarray}
Considering the renormalization parameters of the usual scalar QED model, in which $\phi_B=\sqrt{Z_2}\phi$, $A_{\mu}=\sqrt{Z_3}A_{\mu}$, $m_B^2=Z_mZ_2^{-1}m^2$, $e_B=Z_1Z_3^{-1/2}Z_2^{-1}e$ and $\lambda_B=Z_4Z_2^{-2}\lambda$, we are in a position to renormalize the LV tensors $(\hat{k}_a)^{\mu}$ and $(\hat{k}_c)^{\mu\nu}$. Then a straightforward calculation gives us

\begin{eqnarray}
 (\hat{k}_a^{\mu})_B&=&Z_kZ_2^{-1}(\hat{k}_a^{\mu}),\\
 (\hat{k}_c^{\mu\nu})_B&=&Z_{kk}Z_2^{-1}(\hat{k}_c^{\mu\nu}),
\end{eqnarray}
so that we can rewrite a bare Lagrangian free of divergences as
\begin{eqnarray}
 \nonumber\mathcal{L}_B&=&G^{\mu\nu}_B(D_{\mu B}\phi_B)^*(D_{\nu B}\phi)-m^2_B\phi_B^*\phi_B-\frac{1}{4}F^{\mu\nu}_BF_{\mu\nu B}+\\
 \nonumber&&-\frac{i}{2}[\phi_B^*(\hat{k}_{a})^{\mu}_B D_{\mu B}\phi_B-\phi_B(\hat{k}_{a})^{\mu}_B(D_{\mu B}\phi_B)^*]+\frac{\lambda_B}{4}(\phi_B^*\phi_B)^2,\\
\end{eqnarray}
where $G^{\mu\nu}_B=g^{\mu\nu}+(\hat{k}_c)^{\mu\nu}_B$, $F^{\mu\nu}_B=\partial^{\mu}A^{\nu}_B-\partial^{\nu}A^{\mu}_B$ and $D_{\mu B}=\partial_{\mu}-ieA_{\mu B}$.

\section{Final Remarks}

In this paper we studied the Lorentz-Violating extension of scalar sector recently proposed by Kostelecky \cite{Edwards:2018lsn}. By coupling the scalar field with the gauge field we focused our attention on the LV extension of the scalar QED. We discussed some features of the model and extracted the Feynman rules. We also calculated the one-loop radiative corrections for the photon and scalar propagators, as well as the corrections for the scalar-scalar-photon vertex, scalar-scalar-photon-photon vertex, and the scalar-scalar-scalar-scalar vertex. We reported here only the divergent contributions since we were interested in renormalizing the model.

We would like to highlight here that, despite the Lorentz symmetry violation, the gauge symmetry remains unbroken, and, therefore, the Ward identity for scalar QED follows, as we can see from the radiative corrections of the three-vertex and four-vertex. The non-emergence of a mass term for the photon also follows from Ward's identity for the photon's 2-point function.

The finite contributions were calculated with a symbolic program based on Mathematica, but they were too lengthy to be reported here. We verified the validity of Furry's theorem in the context of the LV-sQED and calculated the renormalized LV tensors. The results we achieved in this paper are a natural development of the Lorentz-Violating extension of scalar QED. The finite contributions will be shown in a future paper aiming to establish bounds on the LV parameters.  

\acknowledgments

The authors would like to thank the professor J. A. Helay\"{e}-Neto for the comments and valuable discussions.

\newpage

\appendix
\begin{center}
{\bf Appendix: Feynman diagrams}
\end{center}
 \begin{figure}[H]
 \begin{center}
 \includegraphics[scale=0.5]{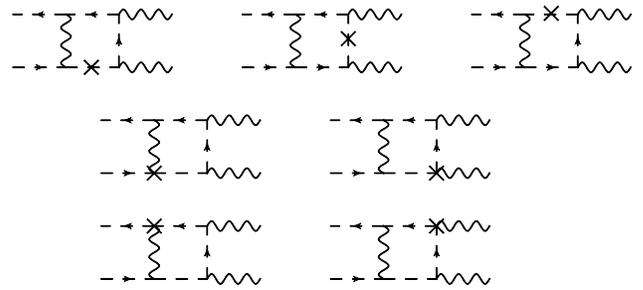}
 \caption{Scalar-scalar-photon-photon vertex correction from $(\hat{k}_a)^{\mu}$ - Contribution 1.}
 \label{d1_ka}
 \end{center}
 \end{figure}
 
 \begin{figure}[H]
 \begin{center}
 \includegraphics[scale=0.5]{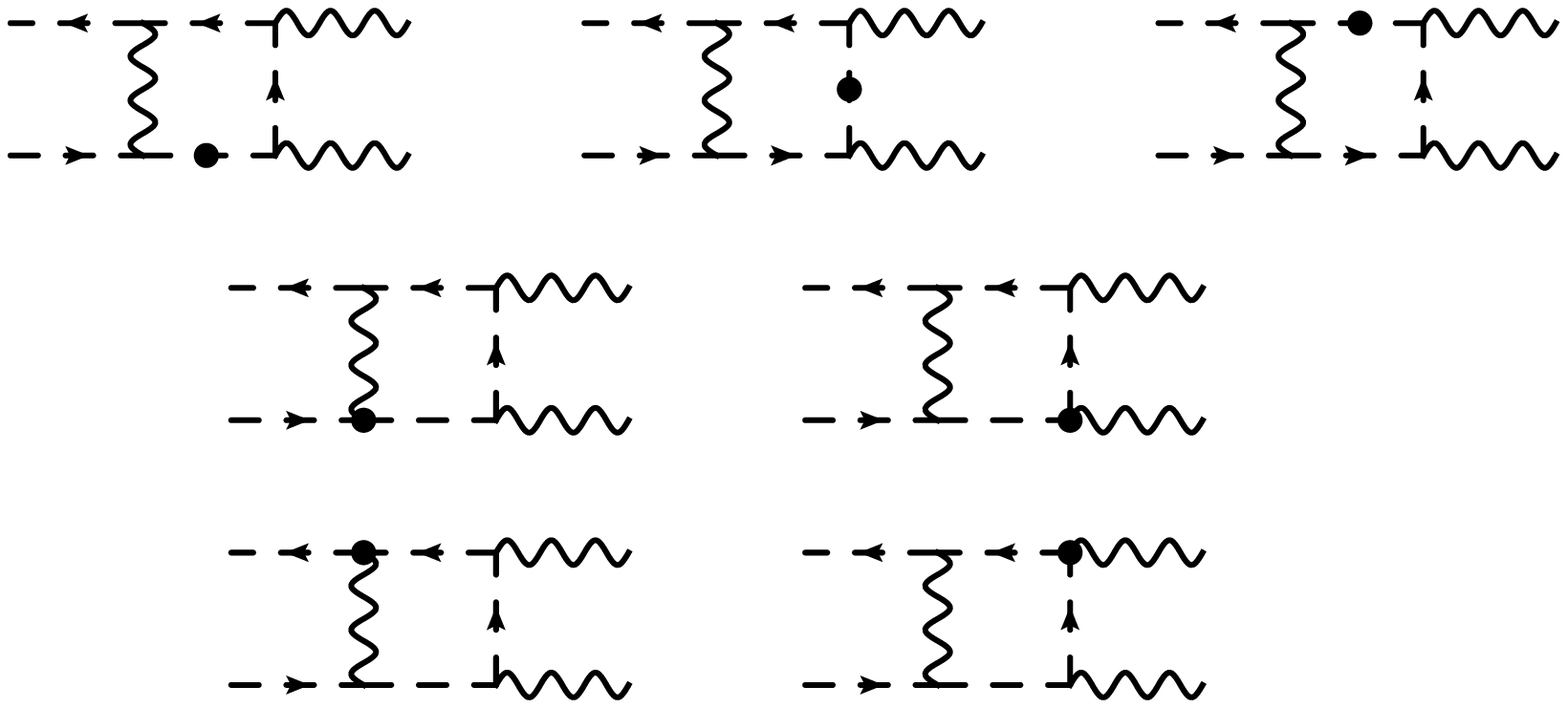}
 \caption{Scalar-scalar-photon-photon vertex correction from $(\hat{k}_c)^{\mu\nu}$ - Contribution 1.}
 \label{d1_kc}
 \end{center}
 \end{figure}
 
 \begin{figure}[H]
 \begin{center}
 \includegraphics[scale=0.5]{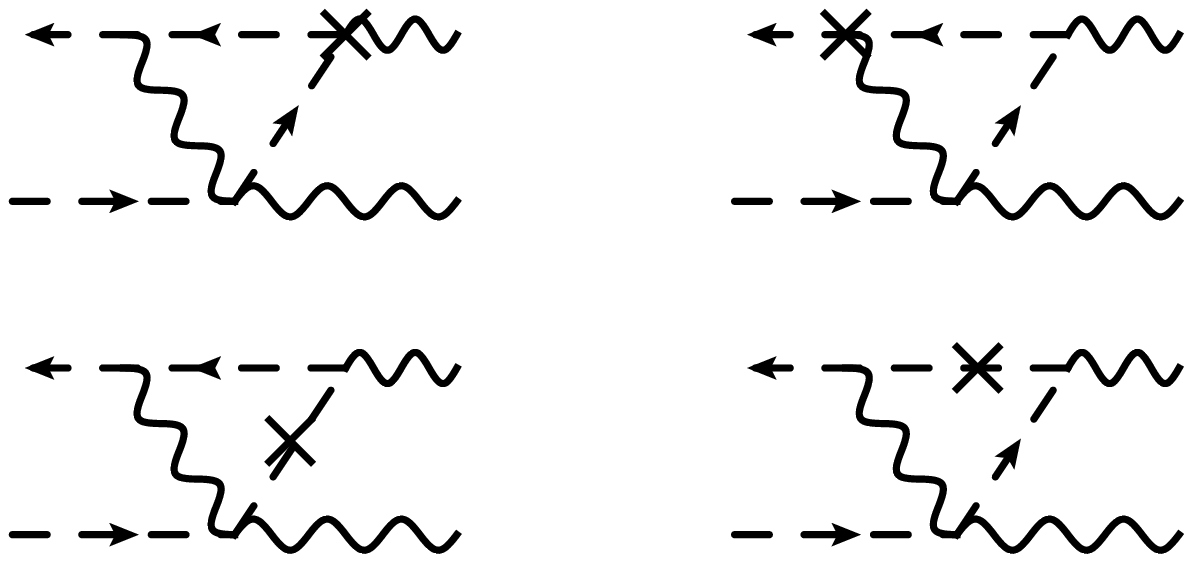}
 \caption{Scalar-scalar-photon-photon vertex correction from $(\hat{k}_a)^{\mu}$ - Contribution 2.}
 \label{d2_ka}
 \end{center}
 \end{figure}
 
 \begin{figure}[H]
 \begin{center}
 \includegraphics[scale=0.5]{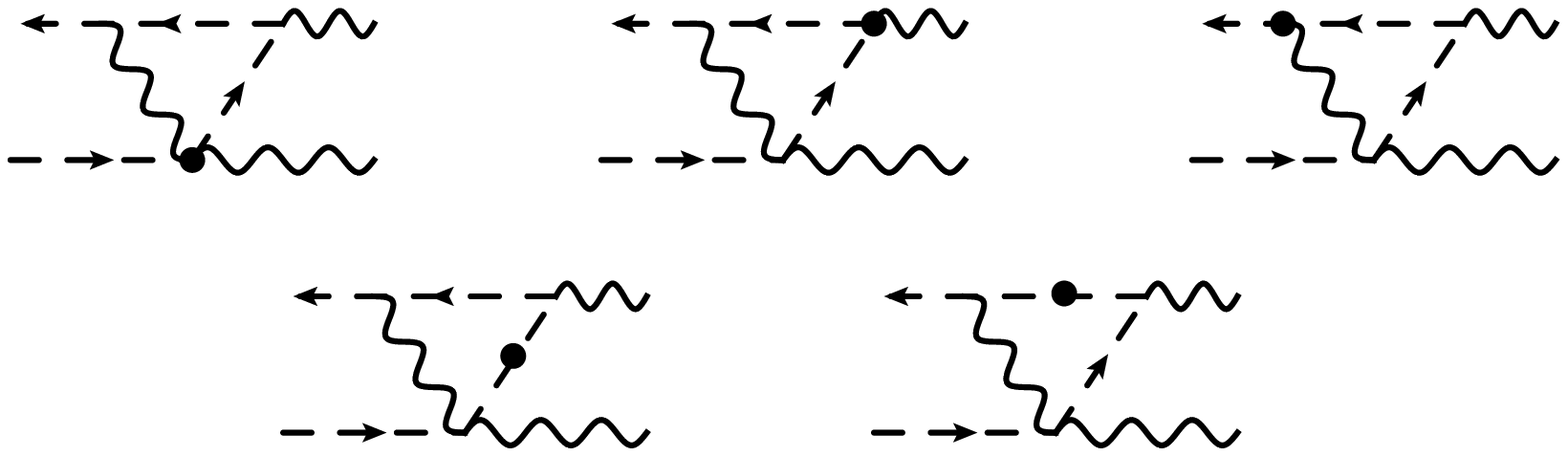}
 \caption{Scalar-scalar-photon-photon vertex correction from $(\hat{k}_c)^{\mu\nu}$ - Contribution 2.}
 \label{d2_kc}
 \end{center}
 \end{figure}
 
 \begin{figure}[H]
 \begin{center}
 \includegraphics[scale=0.5]{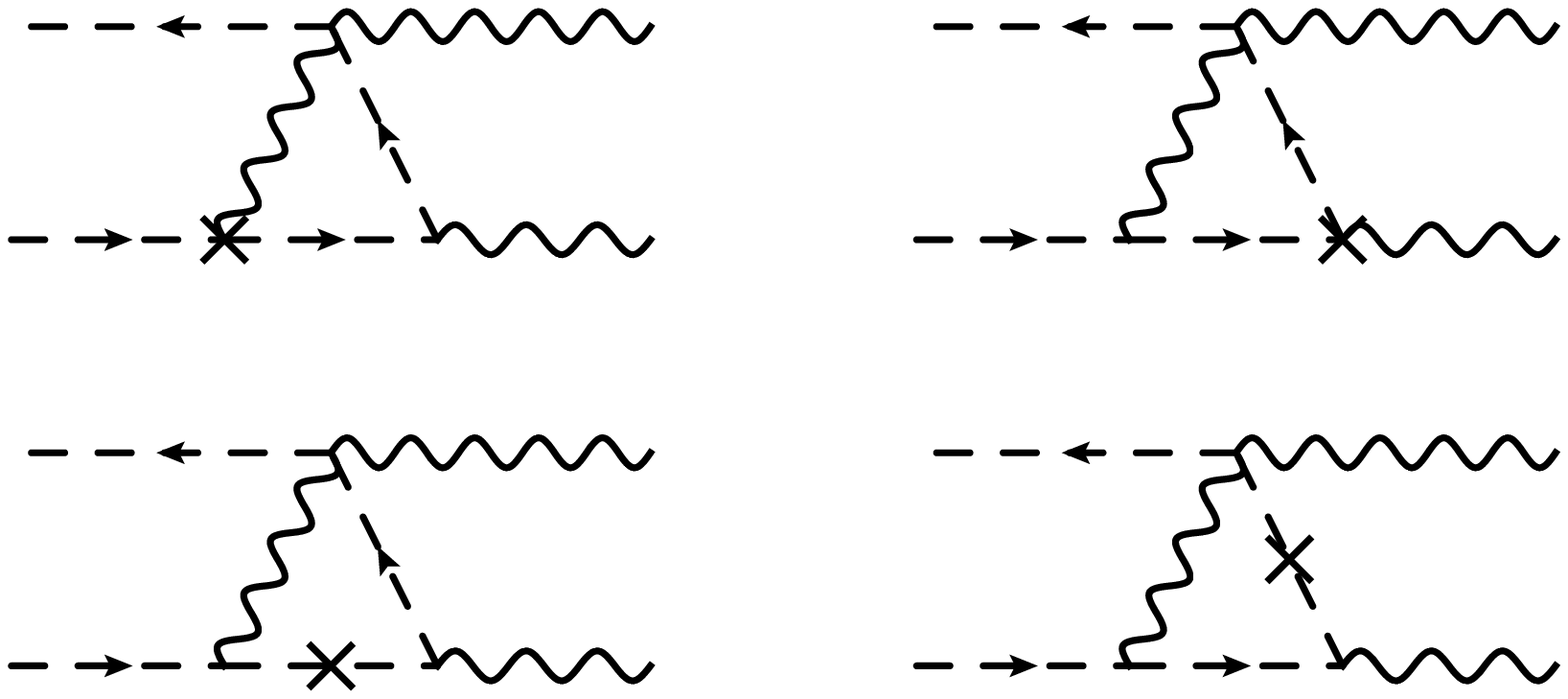}
 \caption{Scalar-scalar-photon-photon vertex correction from $(\hat{k}_a)^{\mu}$ - Contribution 3.}
 \label{d3_ka}
 \end{center}
 \end{figure}

 \begin{figure}[H]
 \begin{center}
 \includegraphics[scale=0.5]{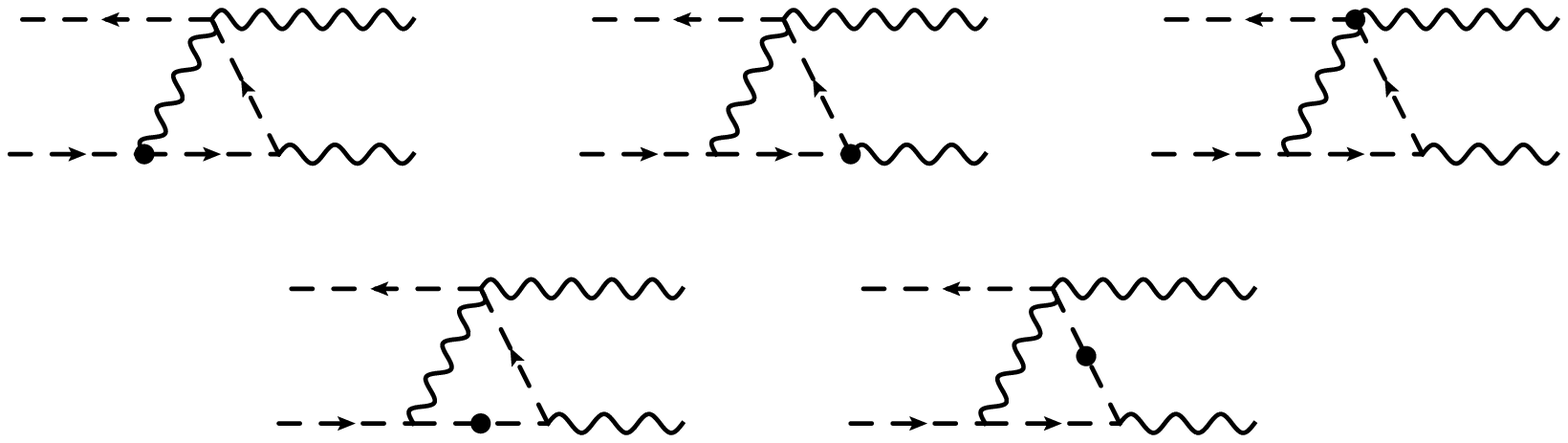}
 \caption{Scalar-scalar-photon-photon vertex correction from $(\hat{k}_c)^{\mu\nu}$ - Contribution 3.}
 \label{d3_kc}
 \end{center}
 \end{figure}
 
 \begin{figure}[H]
 \begin{center}
 \includegraphics[scale=0.5]{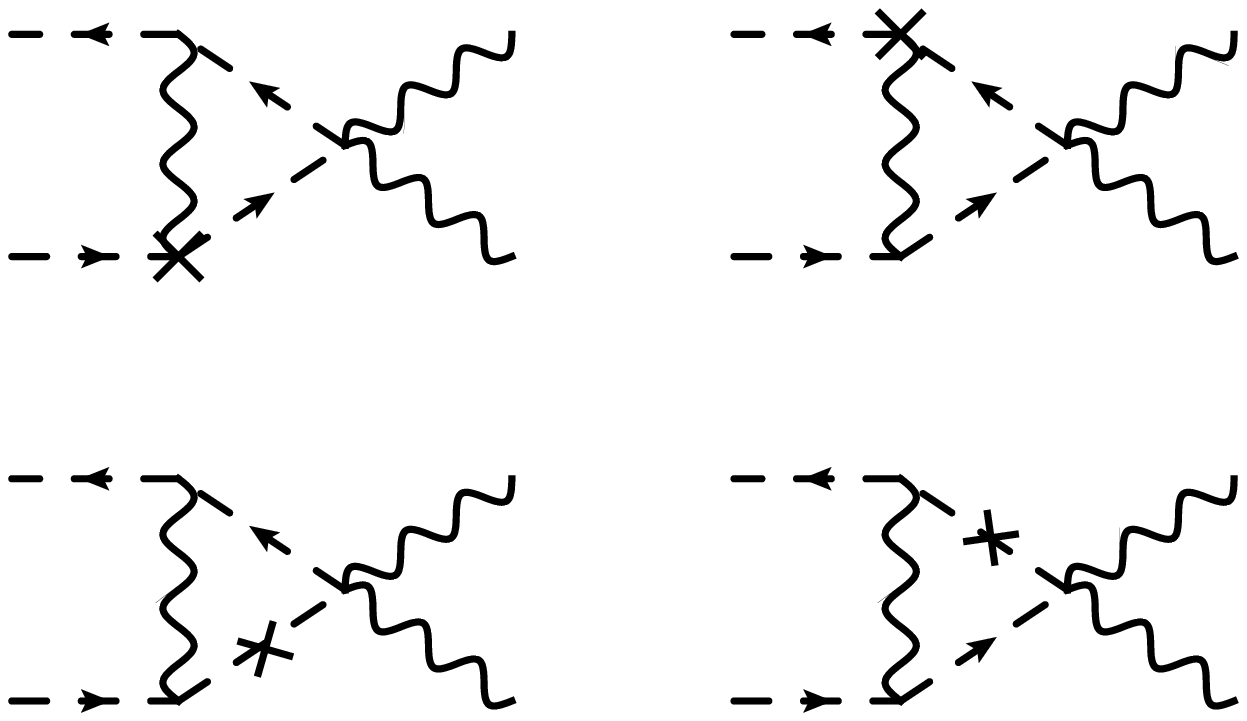}
 \caption{Scalar-scalar-photon-photon vertex correction from $(\hat{k}_a)^{\mu}$ - Contribution 4.}
 \label{d4_ka}
 \end{center}
 \end{figure}
 
 \begin{figure}[H]
 \begin{center}
 \includegraphics[scale=0.5]{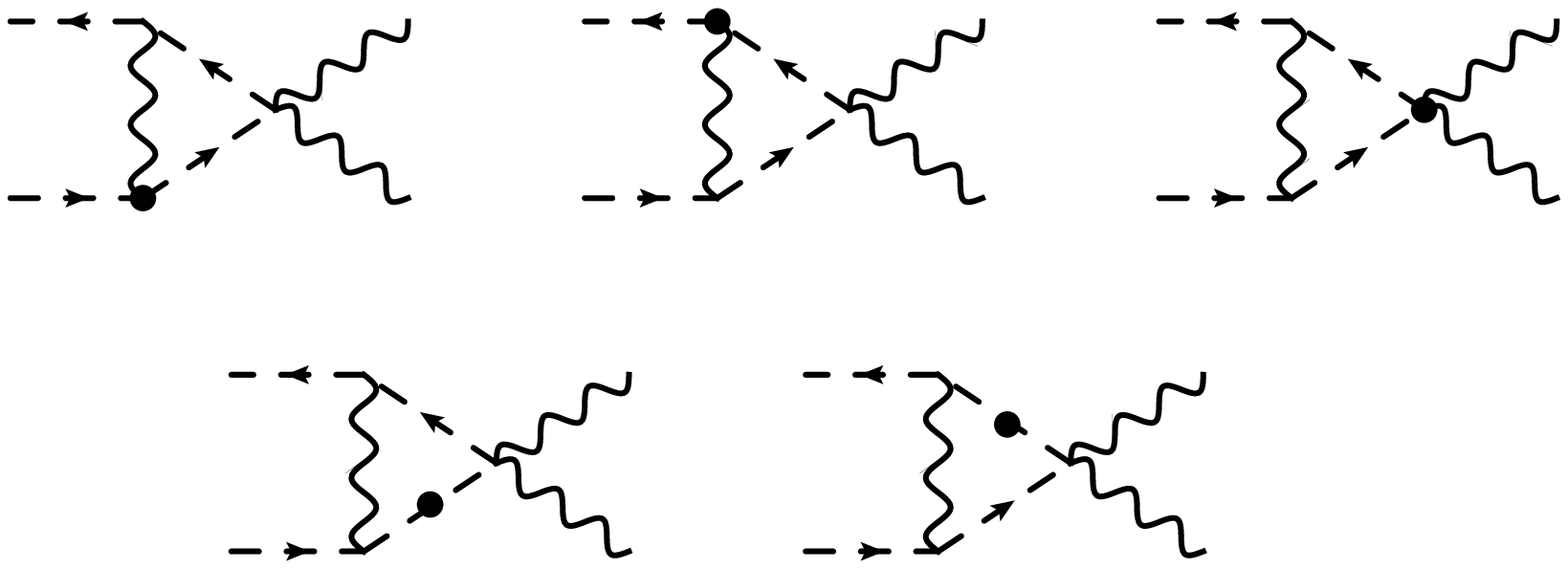}
 \caption{Scalar-scalar-photon-photon vertex correction from $(\hat{k}_c)^{\mu\nu}$ - Contribution 4.}
 \label{d4_kc}
 \end{center}
 \end{figure}
 
 \begin{figure}[H]
 \begin{center}
 \includegraphics[scale=0.5]{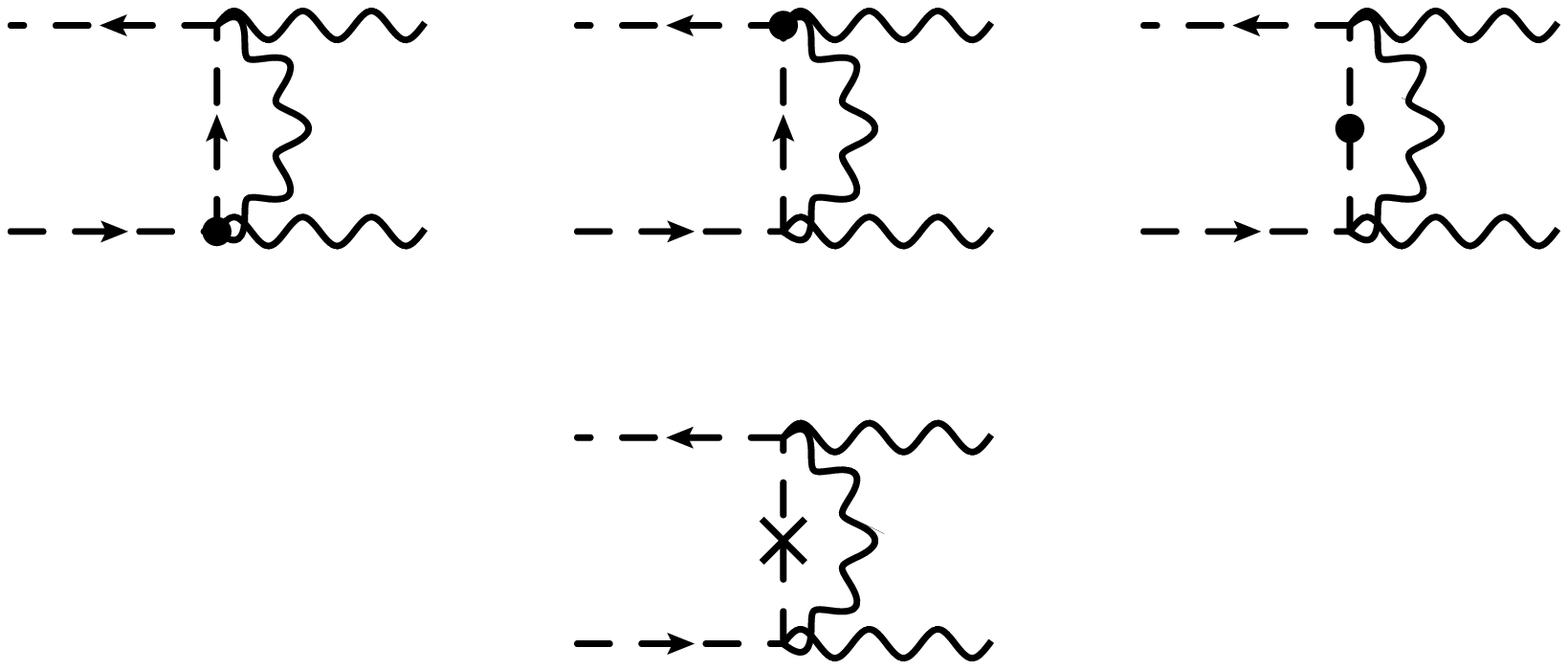}
 \caption{Scalar-scalar-photon-photon vertex correction from $(\hat{k}_a)^{\mu}$ and $(\hat{k}_c)^{\mu\nu}$ - Contribution 5.}
 \label{d5_ka_kc}
 \end{center}
 \end{figure}
 
 \begin{figure}[H]
 \begin{center}
 \includegraphics[scale=0.5]{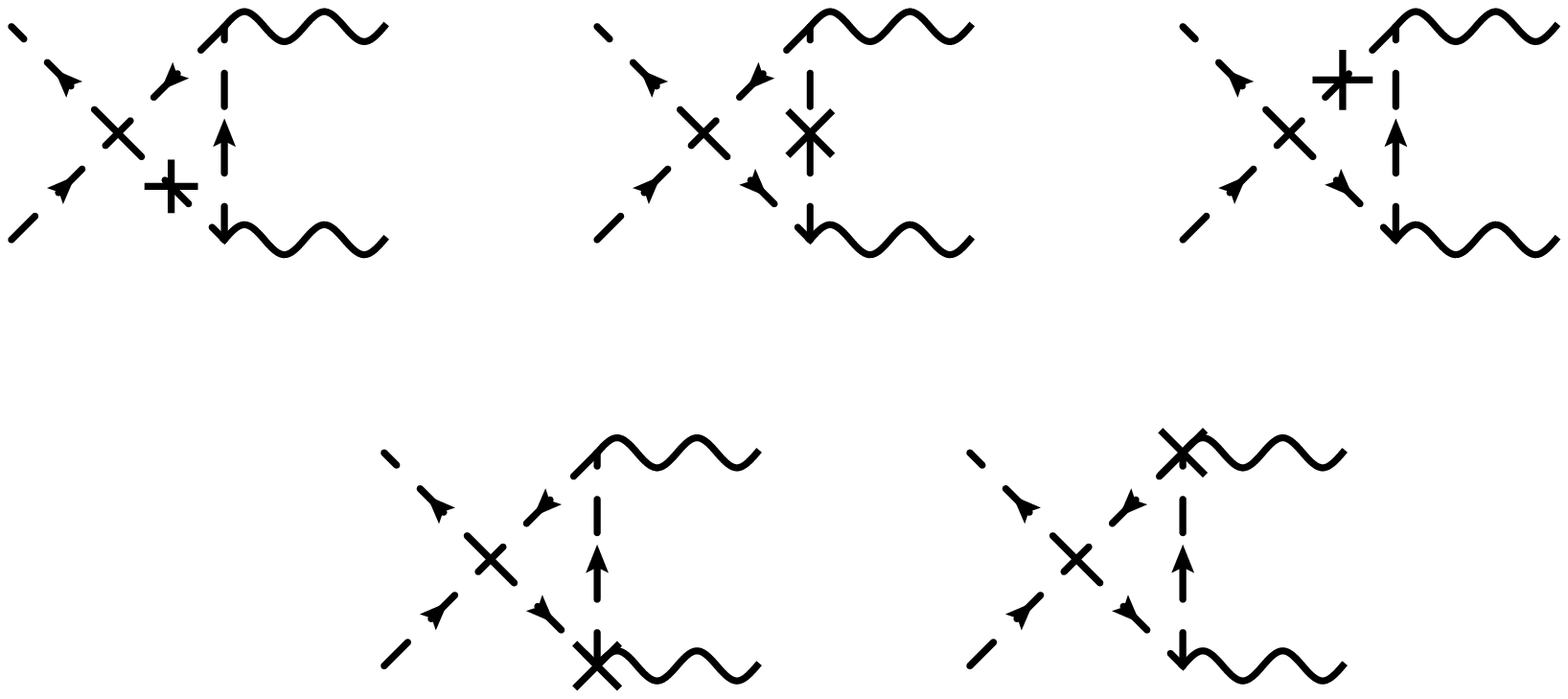}
 \caption{Scalar-scalar-photon-photon vertex correction from $(\hat{k}_a)^{\mu}$ - Contribution 6.}
 \label{d6_ka}
 \end{center}
 \end{figure}
 
 \begin{figure}[H]
 \begin{center}
 \includegraphics[scale=0.5]{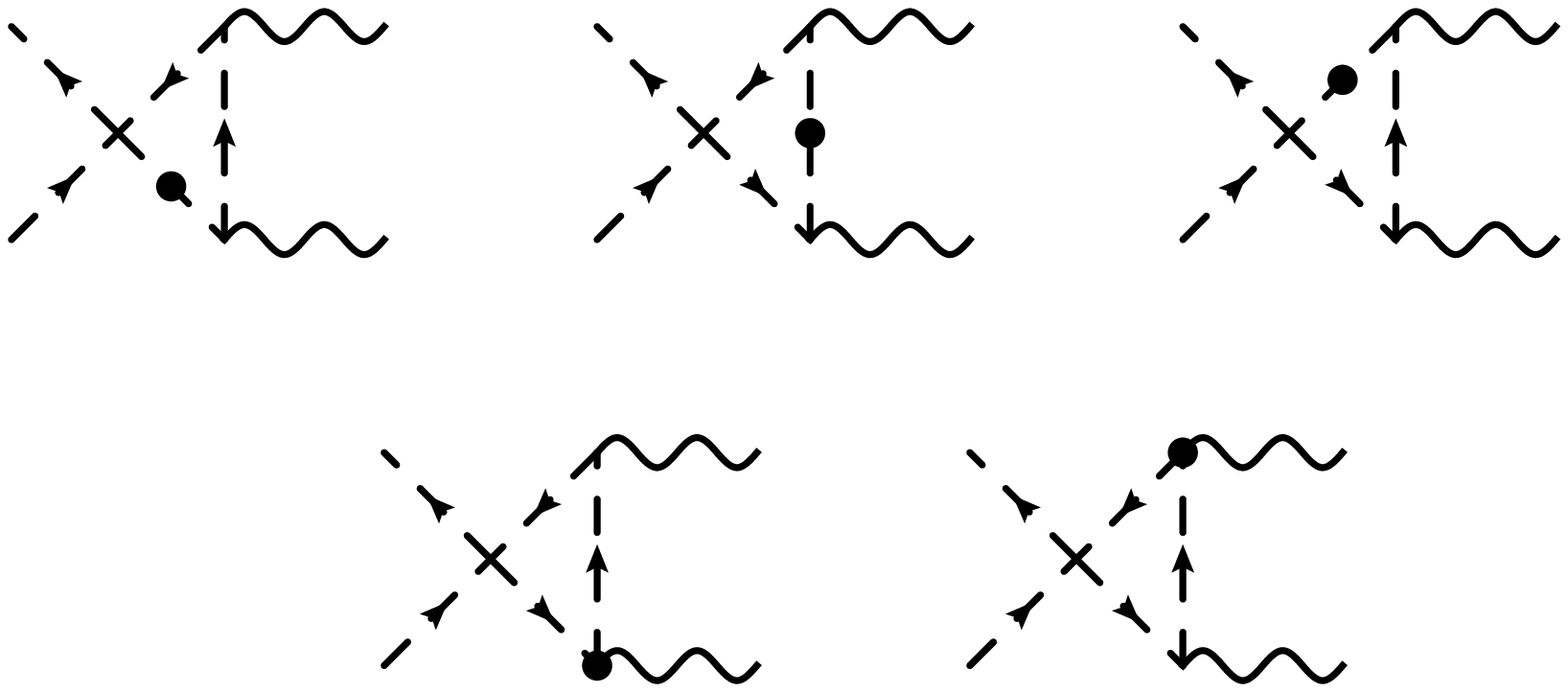}
 \caption{Scalar-scalar-photon-photon vertex correction from $(\hat{k}_c)^{\mu\nu}$ - Contribution 6.}
 \label{d6_kc}
 \end{center}
 \end{figure}
 
 \begin{figure}[H]
 \begin{center}
 \includegraphics[scale=0.5]{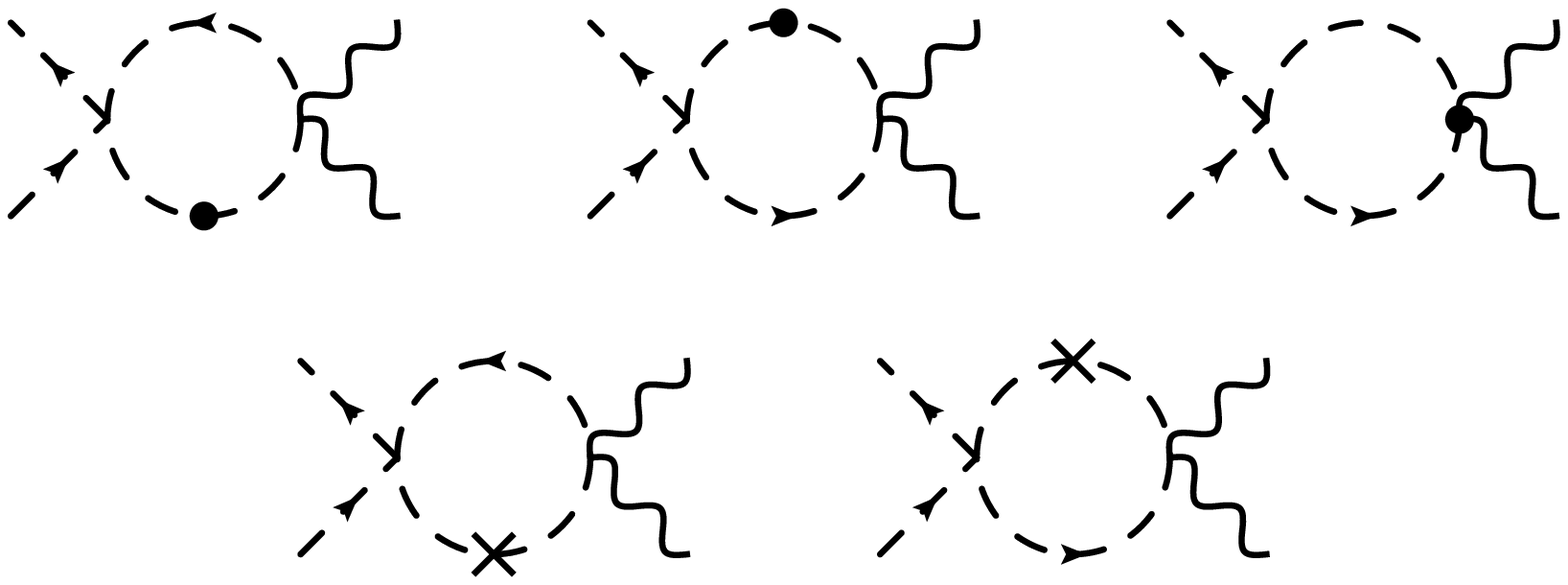}
 \caption{Scalar-scalar-photon-photon vertex correction from $(\hat{k}_a)^{\mu}$ and $(\hat{k}_c)^{\mu\nu}$ - Contribution 7.}
 \label{d7_ka_kc}
 \end{center}
 \end{figure}
 
 \begin{figure}[H]
 \begin{center}
 \includegraphics[scale=0.5]{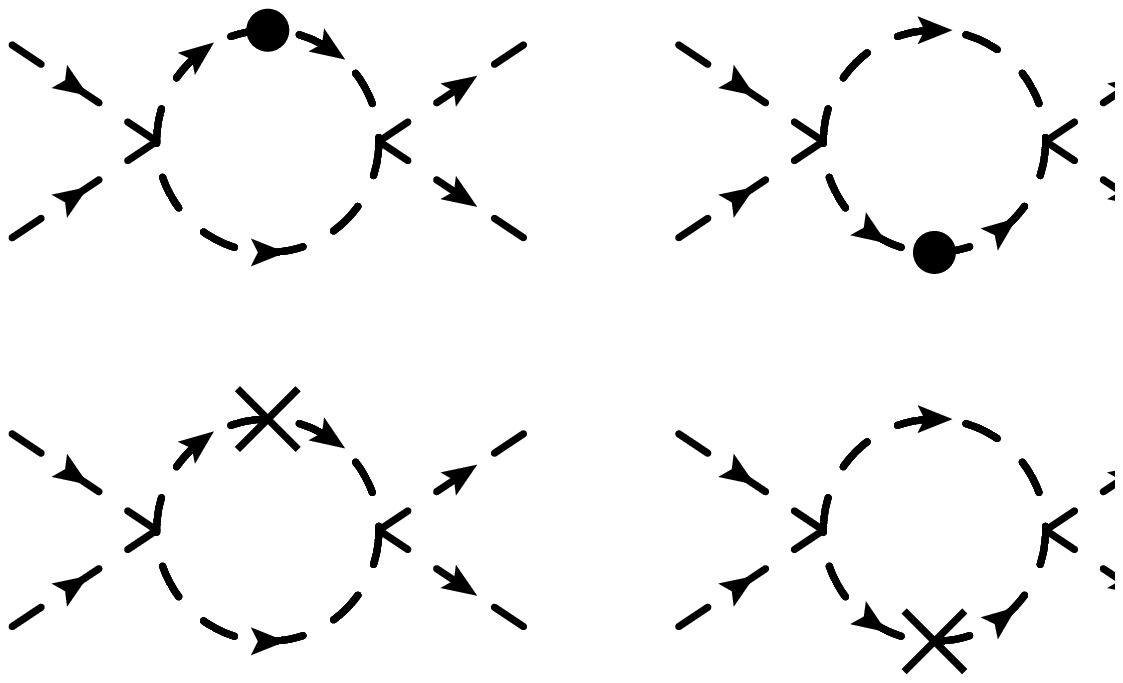}
 \caption{Scalar-scalar-scalar-scalar vertex correction from $(\hat{k}_a)^{\mu}$ and $(\hat{k}_c)^{\mu\nu}$ - Contribution 1.}
 \label{F9}
 \end{center}
 \end{figure}
 
 \begin{figure}[H]
 \begin{center}
 \includegraphics[scale=0.5]{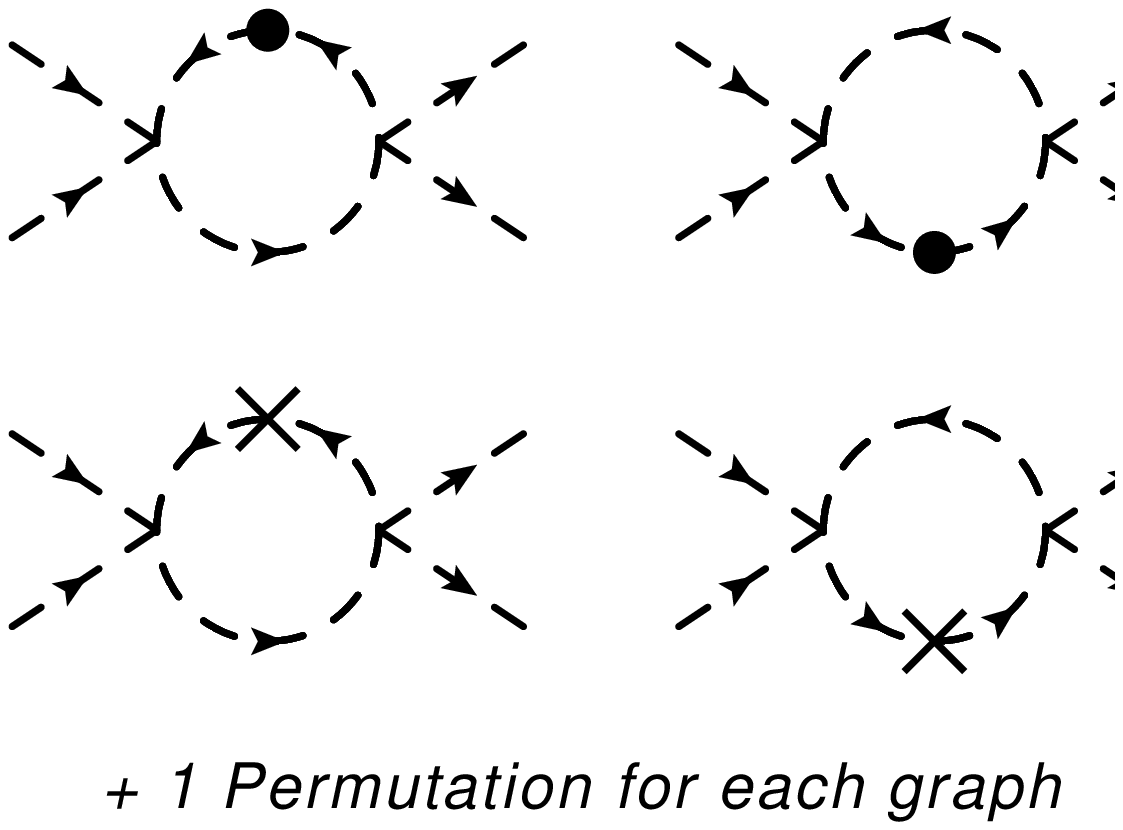}
 \caption{Scalar-scalar-scalar-scalar vertex correction from $(\hat{k}_a)^{\mu}$ and $(\hat{k}_c)^{\mu\nu}$ - Contribution 2.}
 \label{F8}
 \end{center}
 \end{figure}
 
 \begin{figure}[H]
\begin{center}
\includegraphics[scale=0.5]{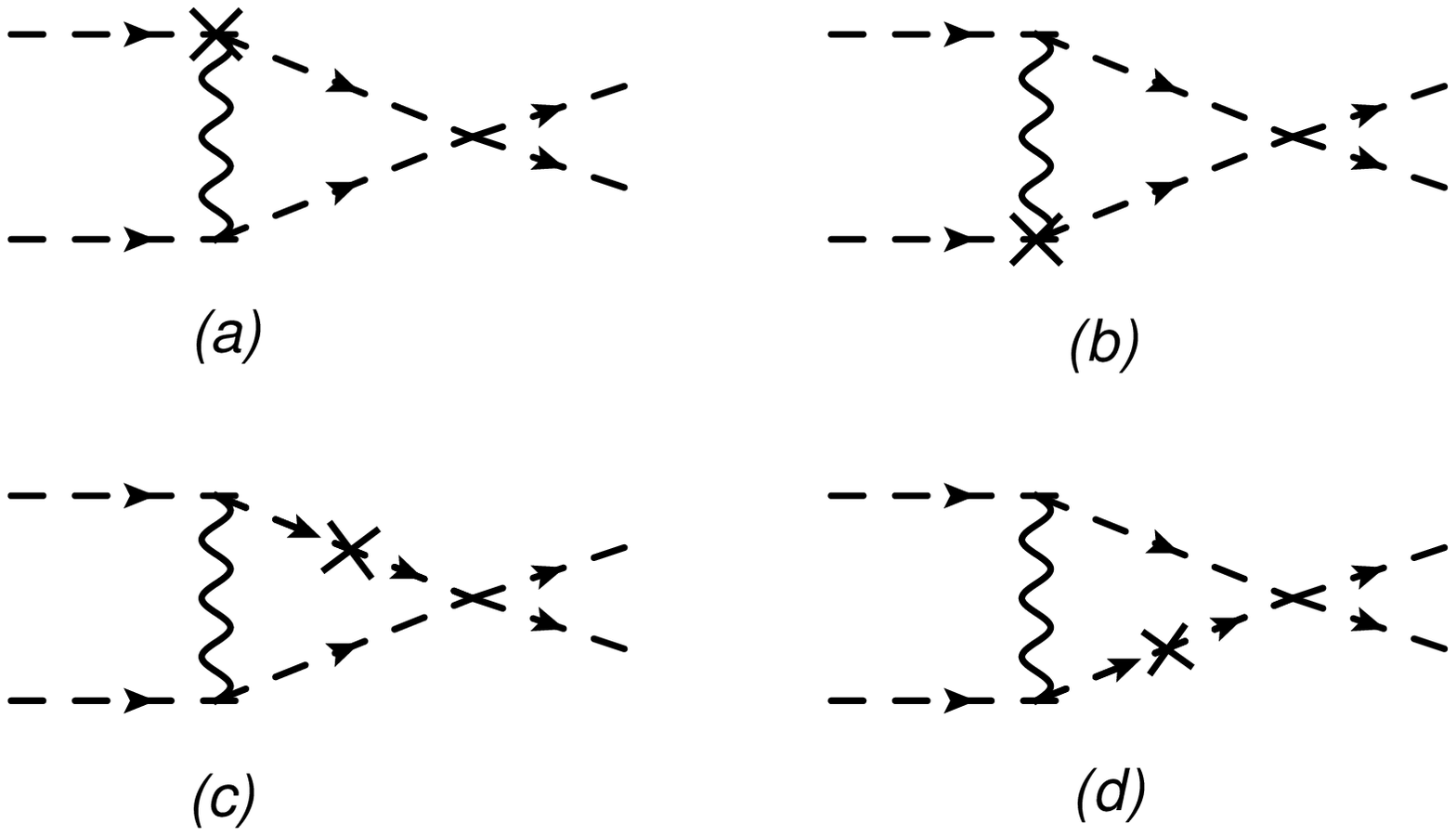}
 \caption{Scalar-scalar-scalar-scalar vertex correction from $(\hat{k}_a)^{\mu}$ - Contribution 3.}
 \label{F4}
 \end{center}
 \end{figure}
 
 \begin{figure}[H]
 \begin{center}
 \includegraphics[scale=0.5]{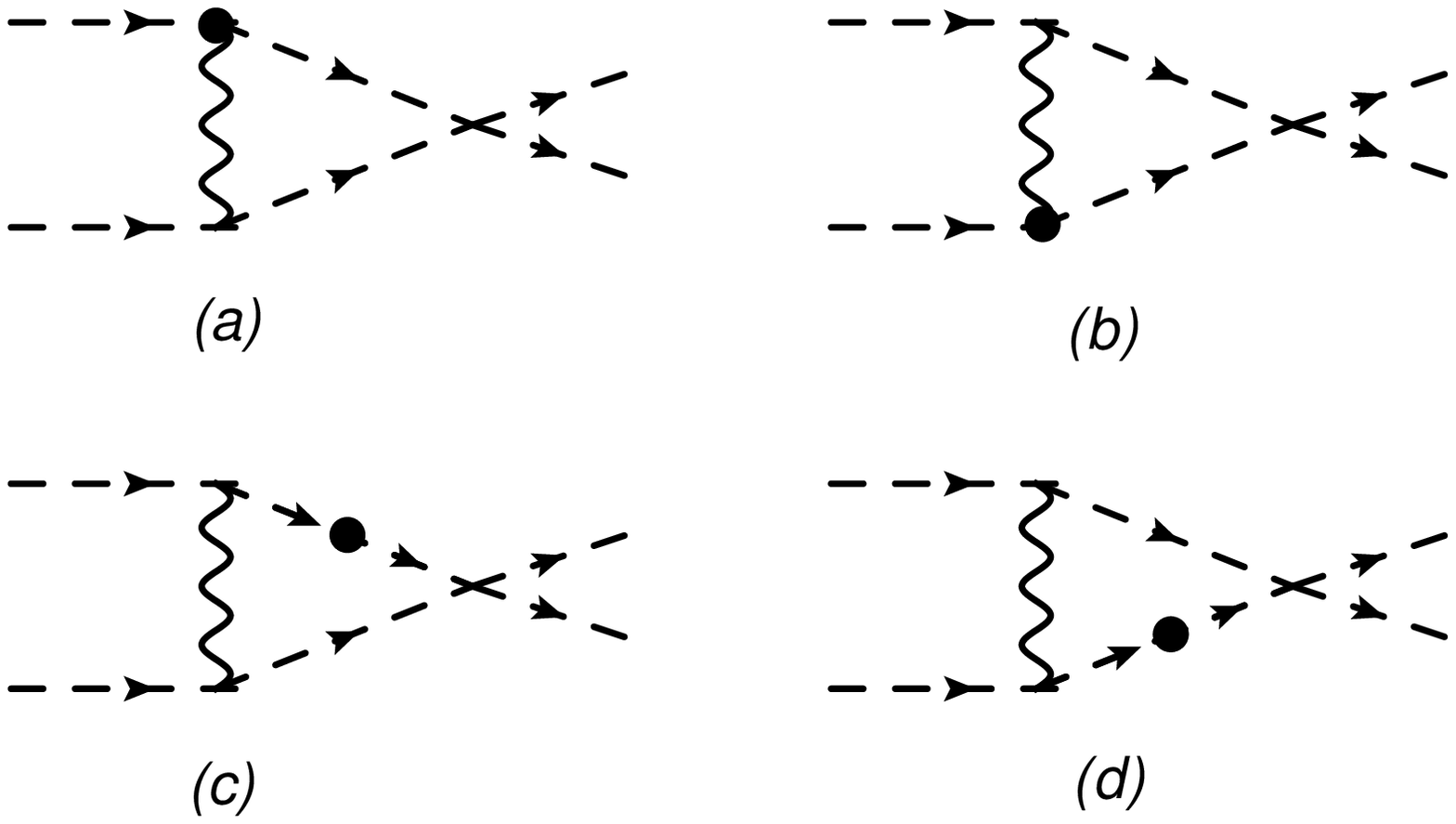}
 \caption{Scalar-scalar-scalar-scalar vertex correction from $(\hat{k}_c)^{\mu\nu}$ - Contribution 4.}
 \label{F41}
 \end{center}
 \end{figure}
 
 \begin{figure}[H]
 \begin{center}
 \includegraphics[scale=0.5]{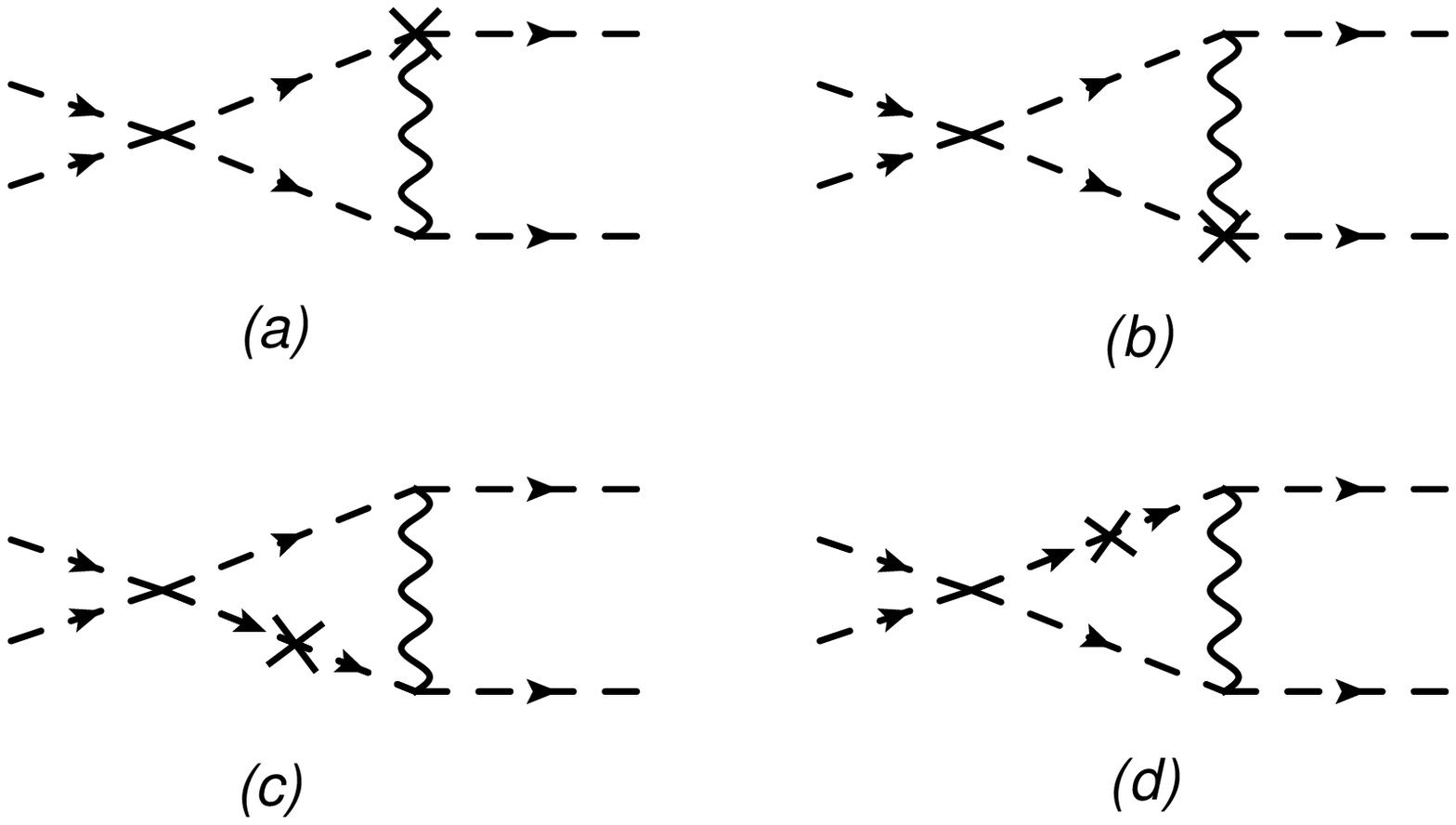}
 \caption{Scalar-scalar-scalar-scalar vertex correction from $(\hat{k}_a)^{\mu}$ - Contribution 5.}
 \label{F5}
 \end{center}
 \end{figure}
 
 \begin{figure}[H]
 \begin{center}
 \includegraphics[scale=0.5]{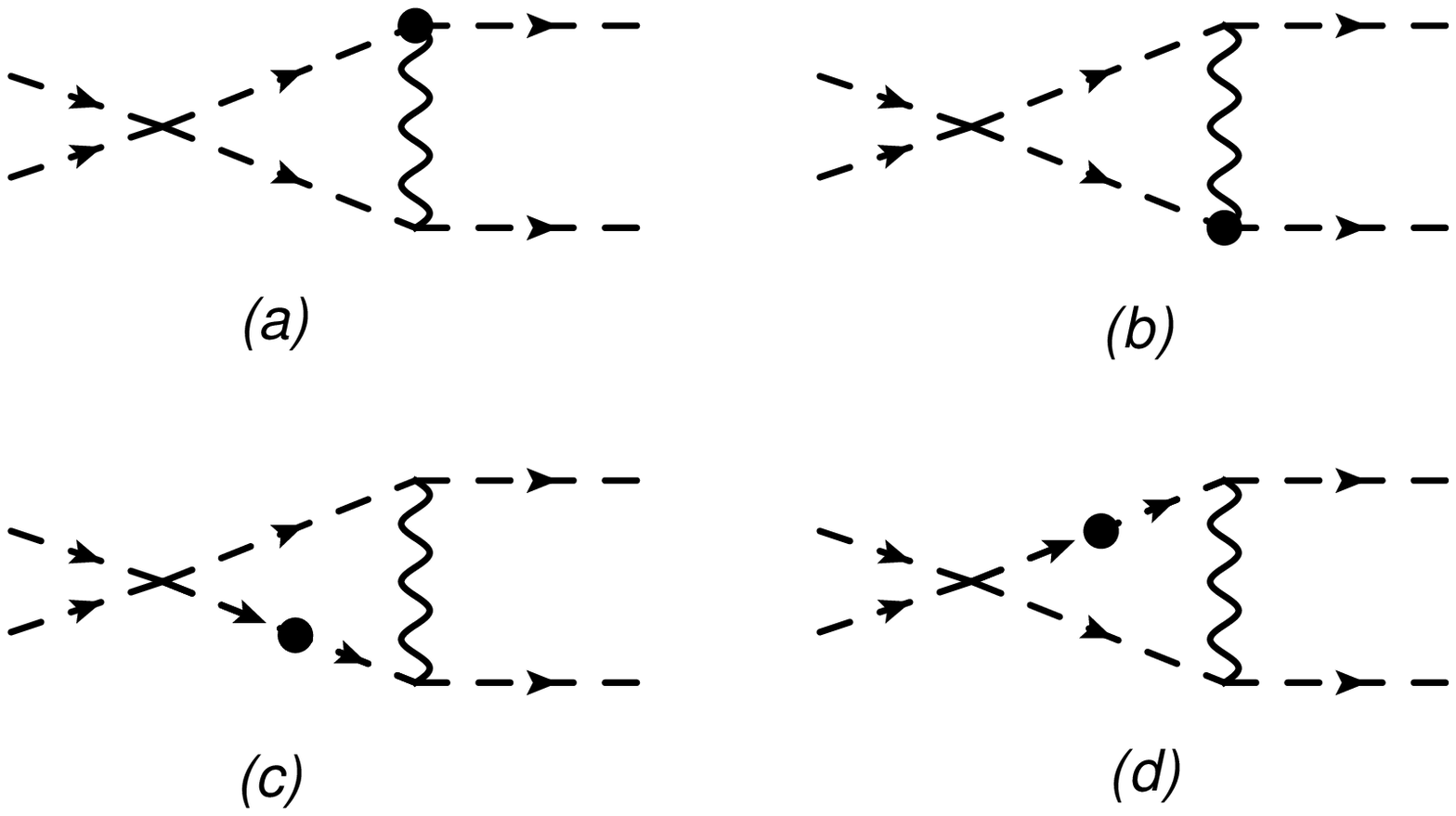}
 \caption{Scalar-scalar-scalar-scalar vertex correction from $(\hat{k}_c)^{\mu\nu}$ - Contribution 6.}
 \label{F51}
 \end{center}
 \end{figure}
 
 \begin{figure}[H]
 \begin{center}
 \includegraphics[scale=0.5]{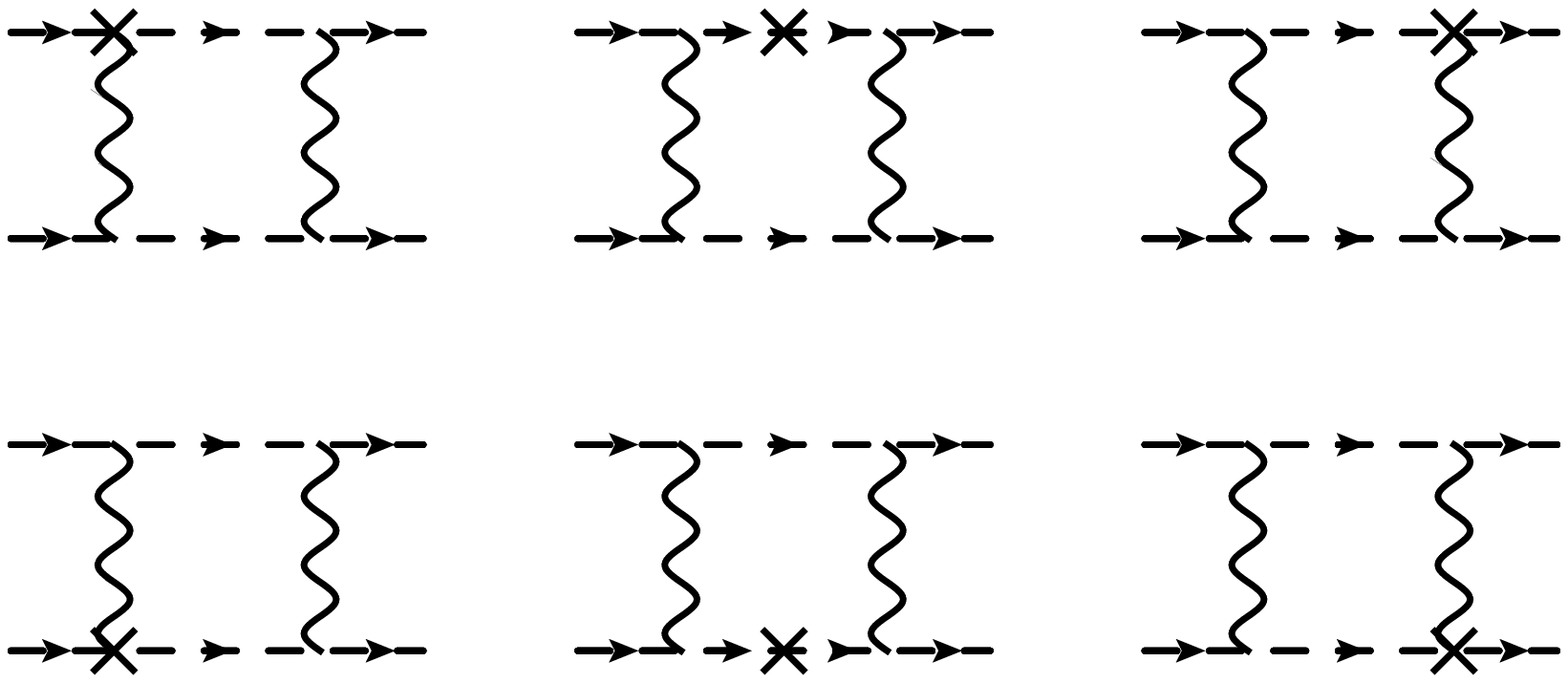}
 \caption{Scalar-scalar-scalar-scalar vertex correction from $(\hat{k}_a)^{\mu}$ - Contribution 7.}
 \label{F1}
 \end{center}
 \end{figure}
 
 \begin{figure}[H]
 \begin{center}
 \includegraphics[scale=0.5]{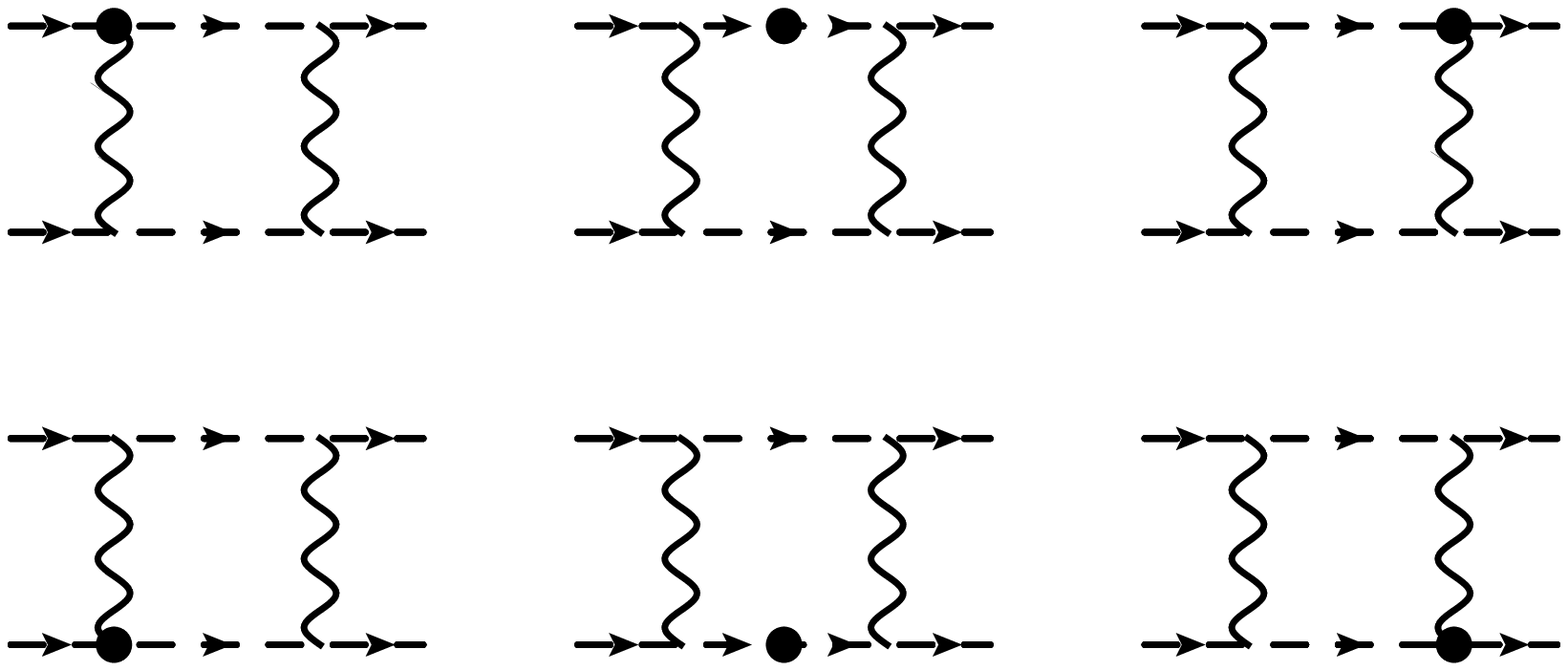}
 \caption{Scalar-scalar-scalar-scalar vertex correction from $(\hat{k}_c)^{\mu\nu}$ - Contribution 8.}
 \label{F11}
 \end{center}
 \end{figure}
 
 \begin{figure}[H]
 \begin{center}
 \includegraphics[scale=0.5]{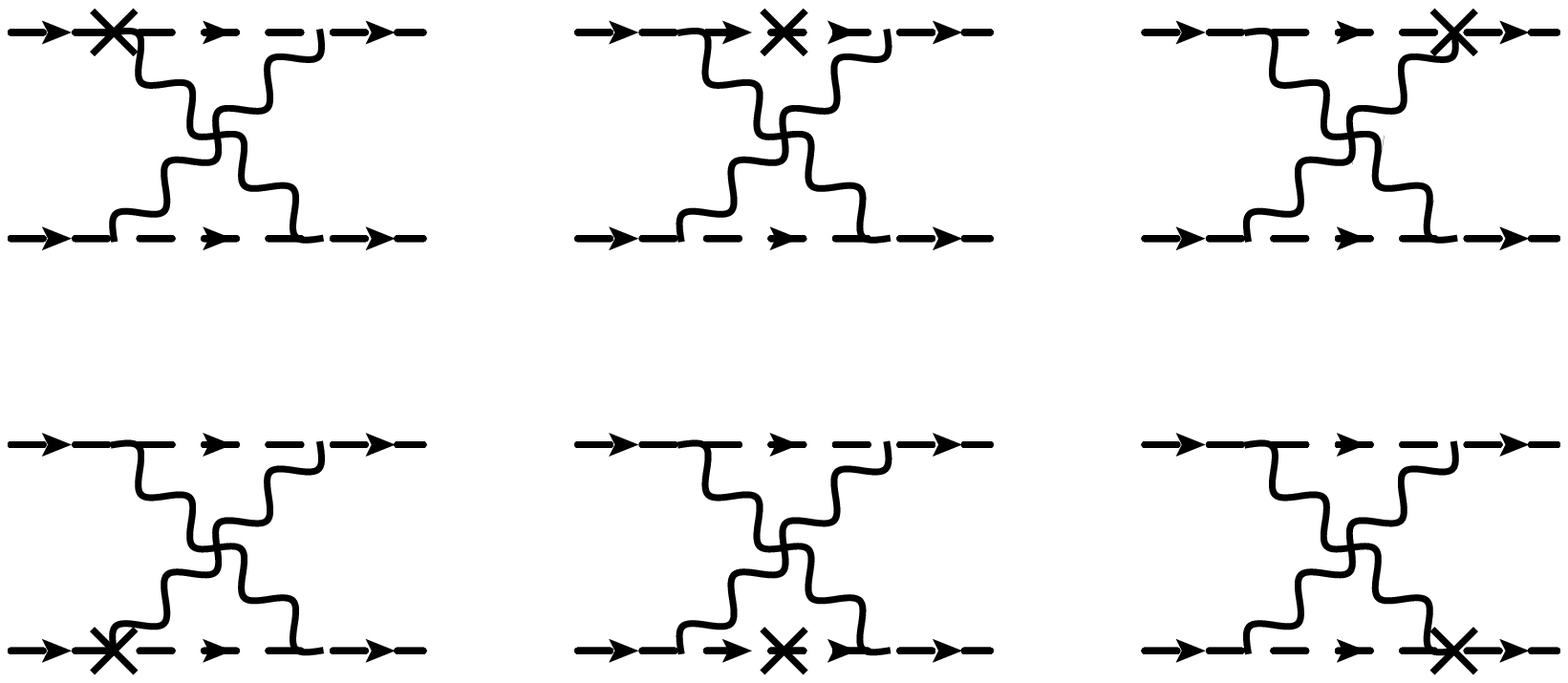}
 \caption{Scalar-scalar-scalar-scalar vertex correction from $(\hat{k}_a)^{\mu}$ - Contribution 9.}
 \label{F2}
 \end{center}
 \end{figure}
 
 \begin{figure}[H]
 \begin{center}
 \includegraphics[scale=0.5]{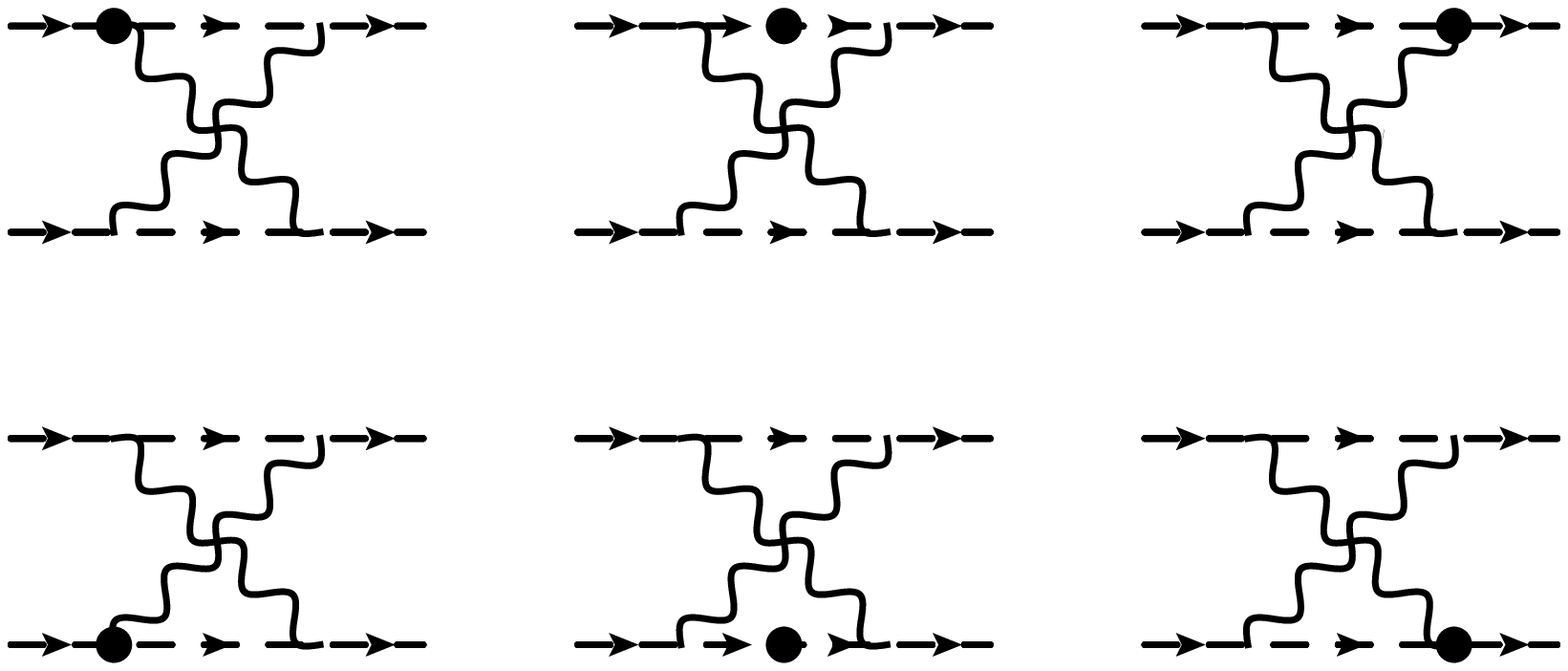}
 \caption{Scalar-scalar-scalar-scalar vertex correction from $(\hat{k}_c)^{\mu\nu}$ - Contribution 10.}
 \label{F21}
 \end{center}
 \end{figure}
 
 \begin{figure}[H]
 \begin{center}
 \includegraphics[scale=0.5]{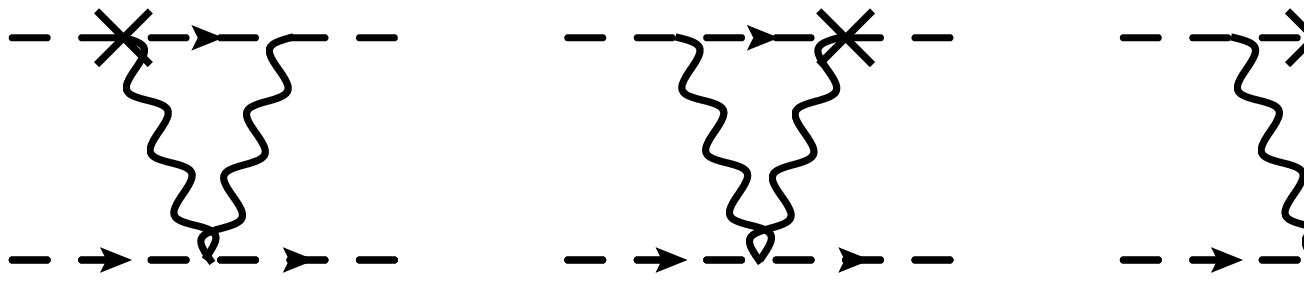}
 \caption{Scalar-scalar-scalar-scalar vertex correction from $(\hat{k}_a)^{\mu}$ - Contribution 11.}
 \label{F3}
 \end{center}
 \end{figure}
 
 \begin{figure}[H]
 \begin{center}
 \includegraphics[scale=0.5]{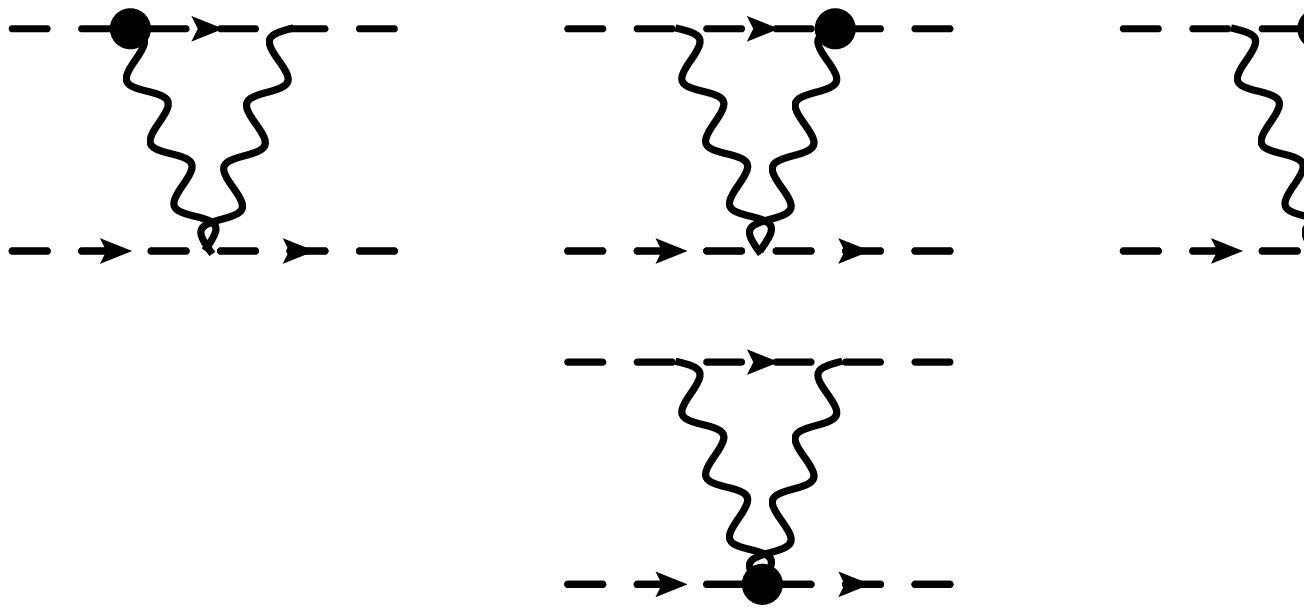}
 \caption{Scalar-scalar-scalar-scalar vertex correction from $(\hat{k}_c)^{\mu\nu}$ - Contribution 11.}
 \label{F31}
 \end{center}
 \end{figure}
 
 \begin{figure}[H]
 \begin{center}
 \includegraphics[scale=0.5]{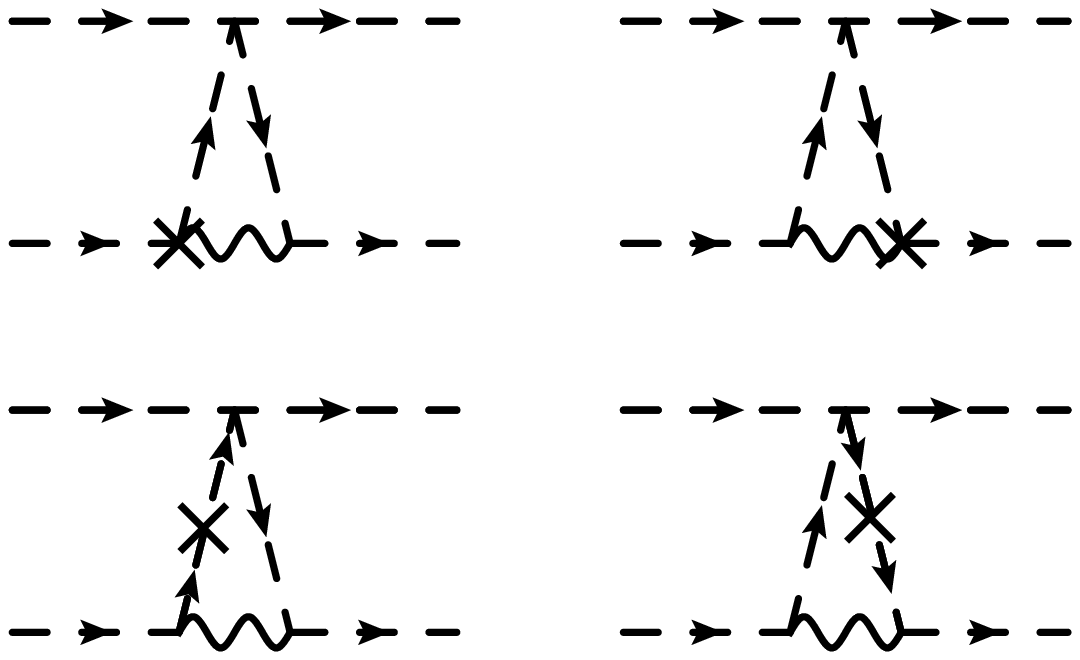}
 \caption{Scalar-scalar-scalar-scalar vertex correction from $(\hat{k}_a)^{\mu}$ - Contribution 12.}
 \label{F6}
 \end{center}
 \end{figure}
 
 \begin{figure}[H]
 \begin{center}
 \includegraphics[scale=0.5]{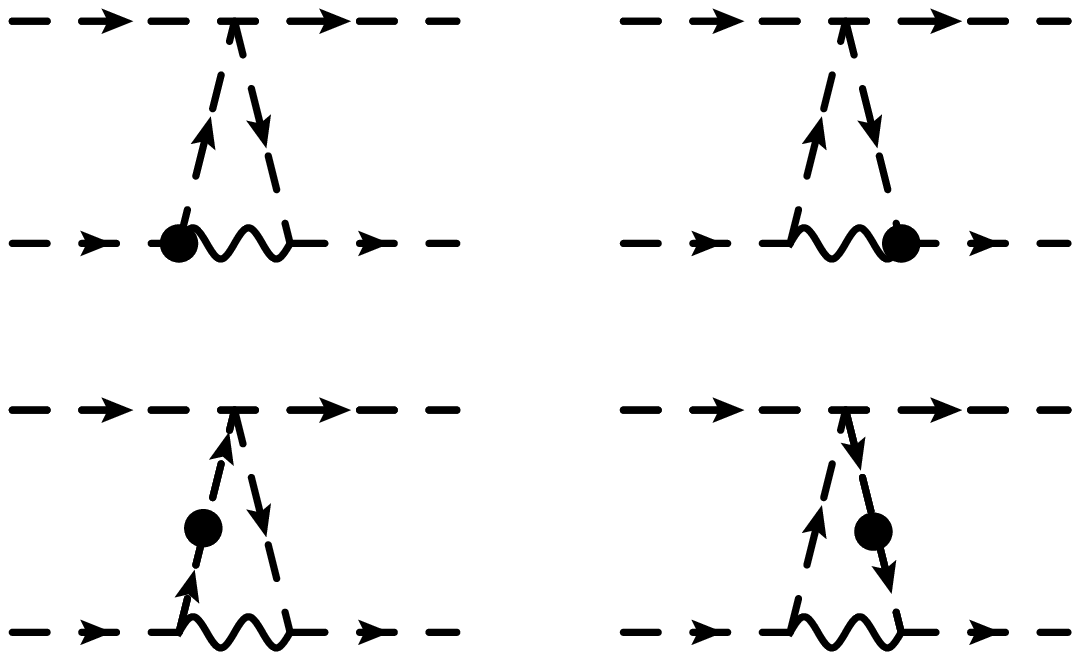}
 \caption{Scalar-scalar-scalar-scalar vertex correction from $(\hat{k}_c)^{\mu\nu}$ - Contribution 13.}
 \label{F61}
 \end{center}
 \end{figure}
 
 \begin{figure}[H]
 \begin{center}
 \includegraphics[scale=0.5]{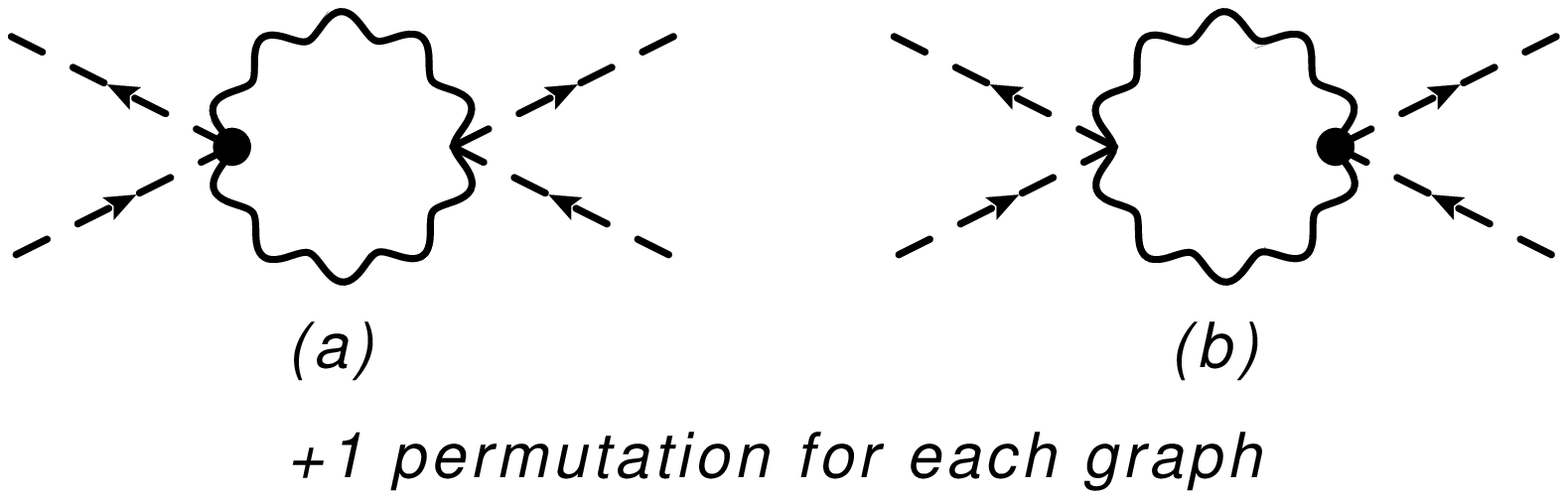}
 \caption{Scalar-scalar-scalar-scalar vertex correction from $(\hat{k}_c)^{\mu\nu}$ - Contribution 14.}
 \label{F7}
 \end{center}
 \end{figure}

\end{document}